\newif\iffullver
\fullvertrue

\documentclass[english]{IEEEtran}
\usepackage[T1]{fontenc}
\usepackage[utf8]{inputenc}
\usepackage{float}
\usepackage{amsthm}
\usepackage{amsmath,amssymb,amsfonts}
\usepackage{graphicx}
\usepackage{xcolor}
\usepackage{subcaption}

\makeatletter

\floatstyle{ruled}
\newfloat{algorithm}{tbp}{loa}
\providecommand{\algorithmname}{Algorithm}
\floatname{algorithm}{\protect\algorithmname}

\theoremstyle{plain}
\newtheorem{thm}{\protect\theoremname}
\theoremstyle{definition}
\newtheorem{defn}[thm]{\protect\definitionname}
\theoremstyle{plain}
\newtheorem{prop}[thm]{\protect\propositionname}
\theoremstyle{remark}
\newtheorem{rem}[thm]{\protect\remarkname}
\theoremstyle{remark}
\newtheorem{eg}{\protect\examplename}
\theoremstyle{plain}
\newtheorem{lem}[thm]{\protect\lemmaname}
\theoremstyle{plain}
\newtheorem{cor}[thm]{\protect\corollaryname}

\usepackage{algorithm,algpseudocode}

\makeatother

\usepackage[english]{babel}
\providecommand{\definitionname}{Definition}
\providecommand{\lemmaname}{Lemma}
\providecommand{\propositionname}{Proposition}
\providecommand{\remarkname}{Remark}
\providecommand{\examplename}{Example}
\providecommand{\theoremname}{Theorem}
\providecommand{\corollaryname}{Corollary}

\usepackage{bbm}

\usepackage{tikz}
\usetikzlibrary{positioning}
\usetikzlibrary{fit}

\begin{document}
\title{Rejection-Sampled Linear Codes for Lossy Compression and Channel Simulation}
\author{Jianguo Zhao and Cheuk Ting Li  \\
Department of Information Engineering, The Chinese University of Hong Kong, Hong Kong, China\\
Email: zhaojg@link.cuhk.edu.hk, ctli@ie.cuhk.edu.hk}
\maketitle
\begin{abstract}
We show that linear codes combined with rejection sampling can yield a capacity-achieving scheme for simulating additive exchangeable noise channels. Specifically, our scheme achieves an amount of communication within $\log e + 1$ bits from the excess functional information lower bound. Hence, it can be used in lossy source coding to achieve the rate-distortion function. We discuss practical implementations based on BCH codes and polar codes. For the simulation of binary symmetric channels, the BCH-based construction with a blocklength of $n = 63$ attains a rate comparable to the PolarSim with $n = 4096$, while significantly reducing the latency. The polar-based construction asymptotically achieves the channel capacity with polynomial average complexity. Furthermore, using the idea from greedy rejection sampling, we propose an algorithm to construct capacity-achieving schemes based on any linear codes. Experiments reveal that our construction can outperform conventional covering codes for lossy source coding with Hamming distortion for a certain range of distortion levels, and performs well even when the blocklength is small (e.g., $n = 24$).
\end{abstract}

\section{Introduction}

\iffullver
Channel simulation 
\iffullver
\cite{bennett2002entanglement,cuff2013distributed,bennet2014reverse} 
\else
\cite{bennett2002entanglement} 
\fi
concerns the setting where the encoder and decoder want to simulate a noisy channel via noiseless communication. One-shot channel simulation, where only one copy of the channel is simulated, has also been studied 
\iffullver
\cite{harsha2010communication,braverman2014public,sfrl_trans}, 
\else
\cite{harsha2010communication,sfrl_trans}, 
\fi
where it was shown that the channel $p_{Y|X}$ with input distribution $p_{X}$ can be simulated using an expected $I(X;Y)+\log(I(X;Y)+1)+O(1)$ number of bits. While most previous channel simulation schemes utilize an unstructured  
\iffullver
approach \cite{harsha2010communication,braverman2014public,sfrl_trans,havasi2019minimal,theis2022algorithms,flamich2023adaptive,flamich2023greedy}, 
\else
approach, 
\fi
many of which have an exponential time complexity, it is also possible to construct schemes using structured codes which result in improved running time efficiency. 
\iffullver
For example, an additive noise channel over $\mathbb{R}^{n}$ can be simulated by combining rejection sampling and lattice codes \cite{ling2024rejection}. Another example 
\else
An example
\fi
is the simulation of $n$ copies of the binary symmetric channels using the polar code, or using a more general linear code with a correction step \cite{sriramu2024fast}. A downside of the approach in \cite{sriramu2024fast} is that the blocklength $n$ must be large (the experiments in \cite{sriramu2024fast} are performed for $n\ge2^{12}$). As a result, this approach can only be applied when the data is large, and requires a significant encoding and decoding delay if the encoding/decoding is performed in a streaming manner.

In this work, we propose efficient channel simulation schemes by integrating linear codes (which we call the inner codes) with rejection sampling \cite{devroye1986nonuniform}.
Our schemes can be applied to exactly simulate any additive exchangeable noise (AXN) channel with finite field vector input, i.e., $\mathbf{X}\in\mathbb{F}_{q}^{n}$, $\mathbf{Y}=\mathbf{X}+\mathbf{Z}$, where $\mathbf{Z}$ has an exchangeable distribution.
Notably, our schemes perform well even at short blocklengths.
We show that:
\begin{itemize}
    \item The rejection-sampled syndrome encoder (RSSE), which combines syndrome coding of linear codes with rejection sampling, can simulate AXN channels.
    For a subclass termed restricted type channels (RTCs), it can achieve an amount of communication within $\log e + 1$ bits from the channel capacity $C=\max_{p_{\mathbf{X}}}I(\mathbf{X};\mathbf{Y})$.
    \item The state-dependent RSSE, which integrates the RSSE with a technique termed RTC-decomposition, can simulate any AXN channel using an amount of communication within $\log e + 1$ bits from the excess functional information lower bound \cite[Proposition 1]{sfrl_trans}
    \begin{equation*}
        \sum_{\mathbf{y} \in \mathbb{F}_q^n} \int_{0}^{1} \mathbb{P}(p_{\mathbf{Y}|\mathbf{X}}(\mathbf{y}|\mathbf{X}) \geq u) \log \mathbb{P}(p_{\mathbf{Y}|\mathbf{X}}(\mathbf{y}|\mathbf{X}) \geq u) \, du,
    \end{equation*}
    which is tighter than channel capacity for finite blocklengths.
    Practical implementations based on BCH codes \cite{boseClassErrorCorrecting1960,hocquenghemCodesCorrecteursDerreurs1959} and polar codes \cite{arikan2009channel,koradaPolarCodesAre2010} are discussed.
    For the simulation of BSCs, the BCH-based construction 
    with a blocklength of $n = 63$ attains a rate comparable to the PolarSim \cite{sriramu2024fast} with $n = 4096$, resulting in a smaller latency while maintaining computational efficiency.
    The polar-based construction asymptotically achieves the channel capacity with $O(n^{3/2}\log n)$ average complexity, enabling scalable high-performance simulation.
    \item The greedy RSSE (GRSSE) is a construction inspired by greedy rejection sampling \cite{harsha2010communication} for simulating any AXN channel.
    As long as the inner code has a sufficiently large distance, the constructed RSSE achieves an amount of communication close to the channel capacity $C$, and hence is capacity-achieving.
\end{itemize}

An application of our scheme is lossy source coding over $\mathbb{F}_{q}^{n}$ with Hamming distortion. By simulating the Hamming ball channel where $\mathbf{Z}$ is uniform over a Hamming ball, we can ensure that the distortion is always below the limit (as in $d$-semifaithful codes \cite{ornstein1990universal,zhang1997redundancy,kontoyiannis2000pointwise}), while attaining the rate-distortion function. Experiment results show that our scheme can outperform conventional linear covering codes \cite{berger2003rate,cohen1985covering} in terms of rate-distortion performance over a certain range of distortion levels. Perhaps surprisingly, unlike conventional linear covering codes which requires a good trade-off between the rate and the covering radius, our scheme only requires the inner linear code to have a large 
distance (instead of a large covering radius). 
As long as the distance is large enough, our scheme can achieve the rate-distortion function, regardless of the rate of the inner code. This makes it significantly easier to find good inner codes for our scheme, giving a constructive approach to attain the rate-distortion function, compared to finding linear covering codes that attain the rate-distortion function where the only known techniques are non-constructive \cite{berger2003rate}. 


Another application is oblivious relaying \cite{sanderovich2008communication,simeone2011codebook,aguerri2019capacity}, where the relay can apply channel simulation to compress the channel output. See Section \ref{subsec:oblivious}.

\fi

\iffullver

\section{Related Works}

\subsection{Channel Simulation}

Steiner \cite{steiner2000towards} first observed that rejection sampling \cite{devroye1986nonuniform} can be used to construct channel simulation schemes.
To maximize the acceptance probability at each iteration, Harsha et al. \cite{harsha2010communication} proposed greedy rejection sampling (GRS) for discrete distributions.
Flamich and Theis \cite{flamich2023adaptive} and Flamich et al. \cite{flamich2023faster} later extended it to general distributions.
Let $L$ denote the number of iterations (samples) required by the GRS with target distribution $P$ and proposal distribution $Q$.
It was shown \cite{harsha2010communication} that $H(L)$ (which reflects the expected communication cost) is bounded between $D_\mathrm{KL}(P\|Q)$ and $D_\mathrm{KL}(P\|Q) + 2\log(D_\mathrm{KL}(P\|Q)+1) + O(1)$.
By introducing the channel simulation divergence $D_\mathrm{CS}$, which generalizes the excess functional information lower bound \cite[Proposition 1]{sfrl_trans}, Goc and Flamich \cite{goc2024causal} tightened both bounds to $D_\mathrm{CS}(P\|Q) \geq D_\mathrm{KL}(P\|Q)$ and $D_\mathrm{CS}(P\|Q) + \log(e+1) \leq D_\mathrm{KL}(P\|Q) + \log(D_\mathrm{KL}(P\|Q)+1) + \log(e+1) + 1$, showing their lower bound is tight.
They also revealed \cite{goc2024causal} that for any causal rejection sampling scheme, the expected time complexity $\mathbb{E}[L]$ grows at least exponentially in the R\'{e}nyi $\infty$-divergence $D_\infty(P\|Q)$.
Poisson functional representation (PFR) \cite{sfrl_trans} provides a noncausal sampling scheme for general distributions.
It achieves an expected communication cost upper-bounded by \cite{li2024pointwise} 
\begin{equation} \label{eq:PFR_bound}
    I(X;Y) + \log(I(X;Y)+2) + 3.
\end{equation}

To accelerate the simulation process, Flamich et al. \cite{flamich2023greedy, flamich2023faster, flamich2022fast} proposed partitioning the sample space for more efficient searching.
The sampling schemes discussed so far are unstructured.
For practical implementation, structured codes are often preferable.
Based on the capacity-achieving polar codes \cite{arikan2009channel}, Sriramu et al. \cite{sriramu2024fast} developed a simulation scheme for $n$ copies of symmetric binary-output channels, which is asymptotically rate-optimal, in the sense that the communication rate approaches the channel capacity, with $O(n \log n)$ complexity.
They also proposed using general linear codes with a correction step to simulate BSCs.
Although both schemes approach the capacity for large $n$, their performance at short blocklengths is not guaranteed.

Refer to \cite{li2024channel} for an overview of various channel simulation schemes.

\subsection{Linear Codes for Lossy Source Coding}
Using linear codes for lossy source coding with Hamming distortion has a long history.
It was shown by Goblick \cite{goblick1963coding} that linear codes can achieve the rate-distortion function.
Due to the duality between channel coding and lossy source coding, many lossy source codes are designed based on good channel codes.
The early attempt of Viterbi and Omura \cite{viterbiTrellisEncodingMemoryless1974} showed that using trellis codes and the Viterbi algorithm \cite{viterbiErrorBoundsConvolutional1967} can approach the rate-distortion function as the trellis constraint length increases.
However, the encoding complexity is exponential in the constraint length.
After the low-density parity-check (LDPC) codes \cite{gallager1962low} attracted a lot of attention in channel coding, Matsunaga and Yamamoto \cite{matsunagaCodingTheoremLossy2003} introduced them to lossy source coding.
They showed that the ensemble of LDPC codes with variable degree $O(\log n)$ can attain the rate-distortion function asymptotically.
Combined with the reinforced belief propagation \cite{braunsteinEncodingBlackwellChannel2007}, the LDPC coding scheme yields near optimal performance empirically with time complexity roughly $O(n \log n)$ \cite{braunsteinEfficientLDPCCodes2009,hondaVariableLengthLossy2014}.
Another line of work considers using the dual code of LDPC codes, i.e., low-density generator matrix (LDGM) codes, for lossy source coding.
Wainwright et al. \cite{wainwrightLossySourceCompression2010} showed that the ensemble of check-regular LDGM codes approaches the rate-distortion function and the gap vanishes rapidly as the degree increases.
Under their proposed message-passing algorithm with a decimation process \cite{cilibertiTheoreticalCapacityParity2005}, the LDGM codes yield performance close to the limit with practical complexity.
The spatially coupled LDGM codes have also been applied to lossy source coding \cite{arefApproachingRatedistortionLimit2015}.
Furthermore, polar codes \cite{arikan2009channel} were proven to achieve the rate-distortion function under the successive cancellation encoding with complexity $O(n \log n)$ \cite{koradaPolarCodesAre2010}.

\subsection{Oblivious Relaying\label{subsec:oblivious}}
Another application is the information bottleneck channel, also known as the oblivious relay channel \cite{sanderovich2008communication,simeone2011codebook,aguerri2019capacity}, which concerns the setting where the encoder encodes the message $M$ into an input sequence $\mathbf{X}=(X_1,\ldots,X_n)$, which is sent through a memoryless channel $p_{Y|X}$. A relay observes the output $\mathbf{Y}$ and compresses it into a description $W$ at a rate $R_{\mathrm{relay}}>0$. The decoder observes $W$ and attempts to decode $M$. The relay is required to be oblivious, that is, it cannot access the codebook used by the encoder and the decoder. The capacity is given by the information bottleneck~\cite{tishby1999information} $\max_{p_{U|Y}: I(Y;U)\le R_{\mathrm{relay}}} I(X;U)$ where $X\to Y\to U$ forms a Markov chain \cite{sanderovich2008communication}. For example, consider $p_{Y|X}$ to be a binary symmetric channel $\mathrm{BSC}(\alpha)$. If the relay simulates the channel $p_{U|Y} = \mathrm{BSC}(\beta)$ so as to allow the decoder to recover an auxiliary output $\mathbf{U}$, the overall channel from $\mathbf{X}$ to $\mathbf{U}$ would be $\mathrm{BSC}(\alpha(1-\beta)+(1-\alpha)\beta)$. Hence, the encoder and decoder can utilize standard coding schemes for the BSC on the auxiliary channel from $\mathbf{X}$ to $\mathbf{U}$ in order to achieve the capacity, and no specialized algorithm is needed to accommodate the oblivious relay. In comparison, if the relay utilized lossy source coding to compress $\mathbf{Y}$ (guaranteeing only a small Hamming distance between $\mathbf{Y}$ and $\mathbf{U}$) instead of channel simulation, then it would be difficult to compute the likelihood of $\mathbf{X}$ given $\mathbf{U}$. See Figure~\ref{fig:relay}. Refer to \cite{liu2025nonasymptotic} for the connection between oblivious relaying and channel simulation.


\begin{figure}
\begin{centering}
\includegraphics[scale=0.95]{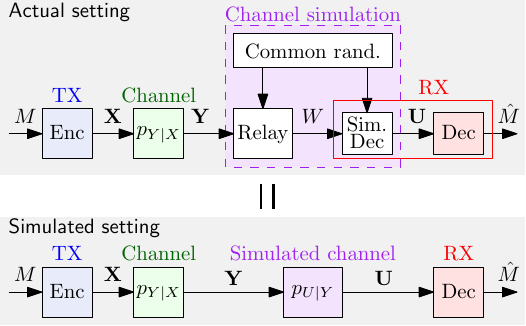}
\par\end{centering}
\caption{\label{fig:relay}Top: An oblivious relay applying channel simulation to simulate a noisy channel $p_{U|Y}$. Bottom: The effect of channel simulation is that the simulated output $\mathbf{U}$ is distributed as if it is the output of $\mathbf{Y}$ passed through another noisy channel $p_{U|Y}$.}
\end{figure}

\subsection*{Notations}

Logarithm is base $2$, and entropy is in bits. Write $\mathbb{N}:=\{1,2,\ldots\}$, $[a:b]:=\{a,\ldots,b\}$, $[n]:=\{1,\ldots,n\}$. For an event $E$, write $\mathbbm{1}\{E\}$ for the indicator function which is $1$ if $E$ occurs, $0$ otherwise. Write $\mathbb{F}_{q}$ for the finite field of order $q$. All vectors are assumed to be row vectors. For a vector $\mathbf{x}\in\mathbb{F}_{q}^{n}$, its Hamming weight is $\mathrm{wt}(\mathbf{x}):=|\{i\in[n]:\,x_{i}\neq0\}|$.

Let $p_X$ denote the probability mass function of a random variable $X$ over $\mathcal{X}$.
The support of $p_X$ is defined as $\mathrm{supp}(p_X) := \{x \in \mathcal{X}:\,p_X(x) > 0\}$.
The uniform distribution over $\mathcal{X}$ is denoted as $\mathrm{Unif}(\mathcal{X})$.
Let $\mathcal{P}_{n}(\mathcal{X})$ denote the set of probability mass functions over $\mathcal{X}$ with entries being multiples of $1/n$. 
For $\mathbf{x} \in \mathcal{X}^{n}$, its type (empirical distribution) is defined as $\mathrm{type}(\mathbf{x}):=(\frac{1}{n}\sum_{i = 1}^{n}\mathbbm{1}\{x_i = x\})_{x \in \mathcal{X}}\in\mathcal{P}_{n}(\mathcal{X})$.
The number of sequences in $\mathcal{X}^n$ of type $\mathbf{t} = (t_1, ..., t_{|\mathcal{X}|}) \in \mathcal{P}_{n}(\mathcal{X})$  is given by the multinominal coefficient 
\begin{align*}
    \binom{n}{n\mathbf{t}} := \frac{n!}{(nt_1)!\cdots(nt_{|\mathcal{X}|})!}.
\end{align*}

\section{The Channel Simulation Setting\label{sec:setting}}

In the setting of (one-shot) channel simulation with common randomness \cite{harsha2010communication}, there is a common randomness $B\sim p_{B}$ available to both the encoder and the decoder. The encoder observes an input $\mathbf{X}$, and sends a description $M=f(B,\mathbf{X})\in\mathcal{C}_{B}$ to the decoder, where $\mathcal{C}_{B}\subseteq\{0,1\}^{*}$ is a prefix-free codebook. The decoder then outputs $\hat{\mathbf{Y}}=g(B,M)$. The objective is to find a scheme $(p_{B},f,g)$ that ensures that $\hat{\mathbf{Y}}$ follows a prescribed conditional distribution $p_{\mathbf{Y}|\mathbf{X}}$ given $\mathbf{X}$ exactly or approximately, while minimizing the amount of communication $\mathbb{E}[|M|]$ (the expected length of $M$). In other words, we are simulating the noisy channel $p_{\mathbf{Y}|\mathbf{X}}$ using the noiseless communication of $M$. See Figure~\ref{fig:cs}.

\begin{figure}
\begin{centering}
\includegraphics[scale=0.95]{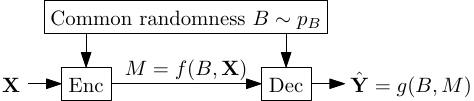}
\par\end{centering}
\caption{\label{fig:cs}One-shot channel simulation.}
\end{figure}

Channel simulation is closely related to lossy source coding \cite{winter2002compression}. To compress $\mathbf{X}$ lossily under the distortion measure $d(\mathbf{x},\mathbf{y})$, we can perform channel simulation for a channel $p_{\mathbf{Y}|\mathbf{X}}$ which satisfies an expected distortion constraint $\mathbb{E}[d(\mathbf{X},\mathbf{Y})]\le D$, or an almost-sure distortion constraint $\mathbb{P}(d(\mathbf{X},\mathbf{Y})\le D)=1$ (e.g., in a $d$-semifaithful code \cite{ornstein1990universal}).

We are interested in simulating a \emph{finite field additive exchangeable noise (AXN) channel}, where the channel input is a finite field vector $\mathbf{X}\in\mathbb{F}_{q}^{n}$, and the channel output is $\mathbf{Y}=\mathbf{X}+\mathbf{Z}$, where $\mathbf{Z}\in\mathbb{F}_{q}^{n}$ is a noise independent of $\mathbf{X}$. We require the entries of $\mathbf{Z}$ to form an exchangeable sequence, i.e., letting $p_{\mathbf{Z}}$ be the probability mass function of $\mathbf{Z}$, we have $p_{\mathbf{Z}}(\mathbf{z})=p_{\mathbf{Z}}(\mathbf{z}')$ if $\mathbf{z}'$ is formed by permuting the entries of $\mathbf{z}$, or equivalently, $p_{\mathbf{Z}}(\mathbf{z})$ depends only on $\mathrm{type}(\mathbf{z})$. In case if the field size is $q=2$, $\mathrm{type}(\mathbf{z})$ is equivalent to the Hamming weight $\mathrm{wt}(\mathbf{z})$, so $p_{\mathbf{Z}}(\mathbf{z})$ only depends on the weight. Note that memoryless additive noise channels, where the entries of $\mathbf{Z}$ are i.i.d., is a special case of AXN channels. We are particularly interested in the following AXN channels:
\begin{itemize}
\item \textbf{Memoryless $q$-ary symmetric channel}, where $\mathbf{Z}=(Z_{i})_{i}$ with $Z_{i}$ being i.i.d. with $p_{Z_{i}}(0)=1-\alpha$, $p_{Z_{i}}(z)=\alpha/(q-1)$ for $z\neq0$. Exactly simulating this channel ensures the expected distortion $\mathbb{E}[\mathrm{wt}(\mathbf{Y}-\mathbf{X})]\le\alpha n$ for Hamming distortion $d(\mathbf{X},\mathbf{Y})=\mathrm{wt}(\mathbf{Y}-\mathbf{X})$.
\item \textbf{Hamming ball channel}, where $\mathbf{Z}\sim\mathrm{Unif}(\{\mathbf{z}:\,\mathrm{wt}(\mathbf{z})\le w\})$ for a fixed $w\in[0:n]$. Exactly simulating this channel ensures the almost-sure distortion constraint $\mathrm{wt}(\mathbf{Y}-\mathbf{X})\le w$.
\item \textbf{Constant type channel}, where $\mathbf{Z}$ is uniform over a type class $\mathbb{F}_q^n\{\mathbf{t}\} := \{\mathbf{z}\in\mathbb{F}_q^n:\,\mathrm{type}(\mathbf{z}) = \mathbf{t}\}$ for a fixed $\mathbf{t} \in \mathcal{P}_n(\mathbb{F}_q)$.
If $q = 2$, the channel is equivalent to a \textbf{constant weight channel}, where $\mathbf{Z}$ is uniform over $\mathbb{F}_2^n\{w\} := \{\mathbf{z} \in \mathbb{F}_2^n:\, \mathrm{wt}(\mathbf{z}) = w\}$ for a fixed $w \in [0:n]$.
Exactly simulating this channel also ensures the almost-sure distortion constraint.
\item \textbf{Restricted type channel (RTC)}, where $\mathbf{Z}$ is uniform over $\mathbb{F}_q^n\{\mathcal{T}\} := \cup_{\mathbf{t} \in \mathcal{T}} \mathbb{F}_q^n\{\mathbf{t}\}$ for a fixed set of types $\mathcal{T} \subseteq \mathcal{P}_n(\mathbb{F}_q)$.
This channel generalizes the Hamming ball and the constant type channels.

\end{itemize}
\medskip{}

\section{Stochastic Syndrome Coder\label{sec:stoc_syn_coder}}

Syndrome encoding and decoding are basic tools for linear codes. In this paper, we slightly generalizes the notion of syndrome coder by allowing a randomized output.

\smallskip{}

\begin{defn} \label{def:syndrome_coder}
An $(n,k)$\emph{-stochastic syndrome coder} is characterized by a pair $(\mathbf{H},\kappa)$, where $\mathbf{H}\in\mathbb{F}_{q}^{(n-k)\times n}$ is a parity-check matrix with full row rank, and $\kappa$ is a conditional probability mass function from $\mathbb{F}_{q}^{n-k}$ to $\mathbb{F}_{q}^{n}$ satisfying that $\kappa(\mathbf{z}|\mathbf{s})=0$ (for $\mathbf{s}\in\mathbb{F}_{q}^{n-k}$, $\mathbf{z}\in\mathbb{F}_{q}^{n}$) if $\mathbf{z}\mathbf{H}^{\top}\neq\mathbf{s}$.
\end{defn}
The concept of (stochastic) syndrome coder is general, and can be found in channel codes and lossy source codes:

\smallskip{}

\begin{itemize}
\item \textbf{Syndrome decoder for channel codes.} The encoder encodes the message $\mathbf{M}\in\mathbb{F}_{q}^{k}$ into the codeword $\mathbf{X}=f(\mathbf{M})\in\mathbb{F}_{q}^{n}$ satisfying $\mathbf{X}\mathbf{H}^{\top}=\mathbf{0}$. The decoder observes $\mathbf{Y}\in\mathbb{F}_{q}^{n}$, generates $\hat{\mathbf{Z}}\in\mathbb{F}_{q}^{n}$ following the conditional distribution $\kappa(\hat{\mathbf{z}}|\mathbf{Y}\mathbf{H}^{\top})$, computes $\hat{\mathbf{X}}:=\mathbf{Y}-\hat{\mathbf{Z}}$, and finds the unique $\hat{\mathbf{M}}\in\mathbb{F}_{q}^{k}$ with $\hat{\mathbf{X}}=f(\hat{\mathbf{M}})$.
\item \textbf{Syndrome encoder for lossy source codes.} The encoder observes $\mathbf{X}\in\mathbb{F}_{q}^{n}$, generates $\hat{\mathbf{Z}}\in\mathbb{F}_{q}^{n}$ following  $\kappa(\hat{\mathbf{z}}|-\mathbf{X}\mathbf{H}^{\top})$, computes $\hat{\mathbf{Y}}:=\mathbf{X}+\hat{\mathbf{Z}}$ (which must satisfy $\hat{\mathbf{Y}}\mathbf{H}^{\top}=\mathbf{0}$), and encodes $\hat{\mathbf{Y}}$ into $\mathbf{M}\in\mathbb{F}_{q}^{k}$. The decoder recovers $\hat{\mathbf{Y}}$ from $\mathbf{M}$ and outputs $\hat{\mathbf{Y}}$.
\end{itemize}
\smallskip{}

A common instance in both settings is the \emph{minimum distance syndrome coder}, where $\kappa$ is taken to be the deterministic mapping which maps the syndrome $\mathbf{s} \in \mathbb{F}_{q}^{n-k}$ to the coset leader ${\arg \min}_{\mathbf{z} \in \mathbb{F}_{q}^{n}:\,\mathbf{z}\mathbf{H}^{\top} = \mathbf{s}}\mathrm{wt}(\mathbf{z})$. 
Another notable instance is the \emph{bounded-distance decoder} for channel coding, where $\kappa$ satisfies $\kappa(\mathbf{z}|\mathbf{z}\mathbf{H}^{\top}) = 1$ for all $\mathbf{z} \in \mathbb{F}_q^n$ with $\mathrm{wt}(\mathbf{z}) \leq \tau$ for a certain error-correction radius $\tau \leq \lfloor (d-1)/2 \rfloor$, and $d$ is the minimum distance of the linear code defined by $\mathbf{H}$.
Note that the conventional bounded-distance decoder does not specify the output of $\kappa$ when the input syndrome $\mathbf{s} \notin \{\mathbf{z}\mathbf{H}^{\top}:\, \mathbf{z} \in \mathbb{F}_q^n, \mathrm{wt}(\mathbf{z}) \leq \tau\}$. 
To match Definition \ref{def:syndrome_coder}, we assume in such cases $\kappa$ outputs an arbitrary $\mathbf{z} \in \mathbb{F}_q^n$ satisfying $\mathbf{z}\mathbf{H}^{\top} = \mathbf{s}$.
For example, if $\mathbf{H}=[\tilde{\mathbf{H}}\,|\,\mathbf{I}_{n-k}]$ is in standard form, $\mathbf{z}$ can be $[\mathbf{m}\,|\,\mathbf{s} - \mathbf{m} \tilde{\mathbf{H}}^{\top}]$ for any $\mathbf{m} \in \mathbb{F}_q^k$.

In this paper, we propose to use the stochastic syndrome coder for channel simulation. Compared to lossy source coding, we include a common randomness $(\boldsymbol{\Pi},\mathbf{B})$ to perform a random permutation and shift, so that the induced noise $\hat{\mathbf{Z}}$ is uniform within any type class (a similar randomization is also present in \cite{sriramu2024fast}).

\smallskip{}

\begin{itemize}
\item \textbf{Stochastic syndrome encoder for channel simulation.} Generate the common randomness $(\boldsymbol{\Pi},\mathbf{B})$ available to the encoder and the decoder, where $\boldsymbol{\Pi}\in\{0,1\}^{n\times n}$ is a uniformly random permutation matrix, and $\mathbf{B}\sim\mathrm{Unif}(\mathbb{F}_{q}^{n-k})$.\footnote{The amount of common randmness needed to encode $(\boldsymbol{\Pi},\mathbf{B})$ is $\log(n!)+(n-k)\log q$ bits, which grows like $O(n\log n)$.} The encoder observes $(\boldsymbol{\Pi},\mathbf{B})$ and $\mathbf{X}\in\mathbb{F}_{q}^{n}$, generates $\hat{\mathbf{V}}\in\mathbb{F}_{q}^{n}$ following the conditional distribution 
\[
\kappa(\hat{\mathbf{v}}\,|\,\mathbf{B}-\mathbf{X}\boldsymbol{\Pi}\mathbf{H}^{\top}),
\]
computes 
\[
\hat{\mathbf{Z}}:=\hat{\mathbf{V}}\boldsymbol{\Pi}^{-1},\;\;\hat{\mathbf{Y}}:=\mathbf{X}+\hat{\mathbf{Z}},
\]
which must satisfy $\hat{\mathbf{Y}}\boldsymbol{\Pi}\mathbf{H}^{\top}=\mathbf{B}$ (there are $q^{k}$ different $\hat{\mathbf{y}}$'s satisfying $\hat{\mathbf{y}}\boldsymbol{\Pi}\mathbf{H}^{\top}=\mathbf{B}$), and encodes $\hat{\mathbf{Y}}$ conditional on $(\boldsymbol{\Pi},\mathbf{B})$ into $\mathbf{M}\in\mathbb{F}_{q}^{k}$. The decoder recovers $\hat{\mathbf{Y}}$ from $\mathbf{M},\boldsymbol{\Pi},\mathbf{B}$ and outputs $\hat{\mathbf{Y}}$. For example, if $\mathbf{H}=[\tilde{\mathbf{H}}\,|\,\mathbf{I}_{n-k}]$ is in standard form, then the encoder can take $\mathbf{M}=(\hat{\mathbf{Y}}\boldsymbol{\Pi})_{1:k}$ to be the first $k$ entries of $\hat{\mathbf{Y}}\boldsymbol{\Pi}$, and the decoder can recover
\begin{equation}
\hat{\mathbf{Y}}=\big[\mathbf{M}\, \big| \,\mathbf{B}-\mathbf{M}\tilde{\mathbf{H}}^{\top}\big]\boldsymbol{\Pi}^{-1}.\label{eq:standard_form}
\end{equation}
\end{itemize}
\smallskip{}

We now verify that the stochastic syndrome encoder simulates an AXN channel.

\smallskip{}

\begin{prop}
\label{prop:sse_noise}The stochastic syndrome encoder simulates an AXN channel, with noise distribution
\[
\mathbb{P}(\hat{\mathbf{Z}}=\hat{\mathbf{z}})=\mathbb{E}_{\boldsymbol{\Pi},\mathbf{S}}\left[\kappa(\hat{\mathbf{z}}\boldsymbol{\Pi}|\mathbf{S})\right],
\]
where $\boldsymbol{\Pi}\in\{0,1\}^{n\times n}$ is a uniformly random permutation matrix, independent of $\mathbf{S}\sim\mathrm{Unif}(\mathbb{F}_{q}^{n-k})$. 
\end{prop}
\begin{IEEEproof}
We have
\begin{align*}
\mathbb{P}(\hat{\mathbf{Z}}=\hat{\mathbf{z}}\,|\,\mathbf{X}=\mathbf{x}) & =\mathbb{E}_{\boldsymbol{\Pi},\mathbf{B}}\left[\kappa(\hat{\mathbf{z}}\boldsymbol{\Pi}\,|\,\mathbf{B}-\mathbf{x}\boldsymbol{\Pi}\mathbf{H}^{\top})\right]\\
 & =\mathbb{E}_{\boldsymbol{\Pi},\mathbf{S}}\left[\kappa(\hat{\mathbf{z}}\boldsymbol{\Pi}|\mathbf{S})\right].
\end{align*}
Due to the random permutation, $\mathbb{P}(\hat{\mathbf{Z}}=\hat{\mathbf{z}})$ only depends on $\mathrm{type}(\hat{\mathbf{z}})$.
\end{IEEEproof}
\smallskip{}

\section{Rejection-Sampled Syndrome Encoder}

A limitation of the stochastic syndrome encoder is that it is a fixed-length scheme, i.e., the length of the description $\mathbf{M}$ is always $k$. A fixed-length scheme can be suitable for approximate channel simulation where the induced noise $\hat{\mathbf{Z}}$ only approximately follows the distribution of the target noise $\mathbf{Z}$ (e.g., i.i.d. Bernoulli for binary symmetric channel), but is generally not suitable for exact channel simulation where $\hat{\mathbf{Z}}$ must have the same distribution as $\mathbf{Z}$ \cite{li2024channel}. In this section, we will describe a class of variable-length schemes which combine the stochastic syndrome encoder with rejection sampling, suitable for exact simulation.

\medskip{}

\begin{defn}
\label{def:rsse}A \emph{rejection-sampled syndrome encoder (RSSE)} is characterized by a sequence of pairs $(\mathbf{H}_{i},\kappa_{i})_{i\in\mathbb{N}}$, where for $i\in\mathbb{N}$, $\mathbf{H}_{i}\in\mathbb{F}_{q}^{(n-k_{i})\times n}$ is a parity-check matrix with full row rank (we allow the $k_{i}$'s to be different), and $\kappa_{i}$ is a conditional probability mass function from $\mathbb{F}_{q}^{n-k_{i}}$ to $\mathbb{F}_{q}^{n}\cup\{\mathrm{e}\}$ satisfying that $\kappa_{i}(\mathbf{z}|\mathbf{s})=0$ (for $\mathbf{s}\in\mathbb{F}_{q}^{n-k_{i}}$, $\mathbf{z}\in\mathbb{F}_{q}^{n}$) if $\mathbf{z}\mathbf{H}_{i}^{\top}\neq\mathbf{s}$.
\end{defn}
\smallskip{}

We call the sequence of codes $(\mathbf{H}_{i})_{i}$ the \emph{inner code sequence}. Definition \ref{def:rsse} describes a general class of schemes instead of one specific scheme. 
Note that each $(\mathbf{H}_{i},\kappa_{i})$ in an RSSE is almost a stochastic syndrome encoder, except that $\kappa_{i}$ is allowed to output a ``rejection symbol'' $\mathrm{e}$ which signals that we should reject the current iteration and proceed to $(\mathbf{H}_{i+1},\kappa_{i+1})$. To perform encoding in the RSSE, the encoder performs the same steps as in the stochastic syndrome encoder in Section \ref{sec:stoc_syn_coder} on $(\mathbf{H}_{i},\kappa_{i})$ for $i=1,2,\ldots$ (each with an independent common randomness $(\boldsymbol{\Pi}_{i},\mathbf{B}_{i})$) until the first time where $\hat{\mathbf{V}}_{i}\neq\mathrm{e}$ (let it be iteration $L\in\mathbb{N}$), and selects $\hat{\mathbf{Z}}=\hat{\mathbf{V}}_{L}\boldsymbol{\Pi}_{L}^{-1}$, $\hat{\mathbf{Y}}=\mathbf{X}+\hat{\mathbf{Z}}$. The encoder then encodes $\hat{\mathbf{Y}}$ conditional on $(\boldsymbol{\Pi}_{L},\mathbf{B}_{L})$ into $\mathbf{M}\in\mathbb{F}_{q}^{k_L}$, and sends $(L,\mathbf{M})$. Practically, the encoder can encode $L$ using the Huffman code \cite{huffman1952method}, Golomb code \cite{golombRunlengthEncodingsCorresp1966} or Elias code \cite{elias1975universal}, and then append $\mathbf{M}\in\mathbb{F}_{q}^{k_L}$ to form the transmitted bit sequence, which is prefix-free since the decoder can recover $L$ and then know the length of $\mathbf{M}$ which is $k_L$. The decoder can then decode $\mathbf{M}$ into $\hat{\mathbf{Y}}$ conditional on $(\boldsymbol{\Pi}_{L},\mathbf{B}_{L})$.
If an optimal prefix-free code is used, the expected communication cost is at most $H(L)+\mathbb{E}[k_L]\log q+1$.

The precise algorithm for RSSE is given in Algorithm \ref{alg:RSSE}, where we assume there are two synchronized random number generators (RNGs) available to the encoder and the decoder which can be used to generate the common randomness $(\boldsymbol{\Pi}_{i},\mathbf{B}_{i})$. In practice, they can initialize two pseudo-random number generators with the same seed.

\begin{algorithm}
\textbf{Procedure} $\textsc{Encode}((\mathbf{H}_{i},\kappa_{i})_{i\in\mathbb{N}},\mathbf{X},\mathfrak{G}):$

\textbf{$\;\;\;\;$Input:} rejection-sampled syndrome encoder $(\mathbf{H}_{i},\kappa_{i})_{i\in\mathbb{N}}$,

\textbf{$\;\;\;\;$$\;\;\;\;$$\;\;\;\;$}input $\mathbf{X}\in\mathbb{F}_{q}^{n}$, RNG $\mathfrak{G}$

\textbf{$\;\;\;\;$Output:} iteration $L\in\mathbb{N}$, description $\mathbf{M}\in\mathbb{F}_{q}^{k_L}$

\smallskip{}

\begin{algorithmic}[1]

\For{$i=1,2,\ldots$}

\State{Generate permutation matrix $\boldsymbol{\Pi}_{i}\in\{0,1\}^{n\times n}$ }

\State{$\qquad$and $\mathbf{B}_{i}\sim\mathrm{Unif}(\mathbb{F}_{q}^{n-k_{i}})$ using $\mathfrak{G}$}

\State{Generate $\hat{\mathbf{V}}_{i}\in\mathbb{F}_{q}^{n}\cup\{\mathrm{e}\}$ using local randomness }

\State{$\qquad$following distribution $\kappa_{i}(\hat{\mathbf{v}}\,|\,\mathbf{B}_{i}-\mathbf{X}\boldsymbol{\Pi}_{i}\mathbf{H}_{i}^{\top})$ }

\If{$\hat{\mathbf{V}}_{i}\neq\mathrm{e}$}

\State{$L \leftarrow i$}

\State{$\hat{\mathbf{Z}}\leftarrow\hat{\mathbf{V}}_{L}\boldsymbol{\Pi}_{L}^{-1}$, $\;$$\hat{\mathbf{Y}}\leftarrow\mathbf{X}+\hat{\mathbf{Z}}$}

\State{Encode $\hat{\mathbf{Y}}$ conditional on $(\boldsymbol{\Pi}_{L},\mathbf{B}_{L})$ into $\mathbf{M}\in\mathbb{F}_{q}^{k_L}$}

\State{$\qquad$(e.g., using (\ref{eq:standard_form}) if $\mathbf{H}_{i}$ is in standard form)}

\State{\Return{$(L,\mathbf{M})$}}

\EndIf

\EndFor

\end{algorithmic}

\bigskip{}

\textbf{Procedure} $\textsc{Decode}((\mathbf{H}_{i},\kappa_{i})_{i\in\mathbb{N}},L,\mathbf{M},\mathfrak{G}):$

\textbf{$\;\;\;\;$Input:} $(\mathbf{H}_{i},\kappa_{i})_{i\in\mathbb{N}}$, $L\in\mathbb{N}$, $\mathbf{M}\in\mathbb{F}_{q}^{k_L}$, RNG $\mathfrak{G}$

\textbf{$\;\;\;\;$Output:} output $\hat{\mathbf{Y}}\in\mathbb{F}_{q}^{n}$

\smallskip{}

\begin{algorithmic}[1]

\For{$i=1,\ldots,L$}

\State{Generate permutation matrix $\boldsymbol{\Pi}_{i}\in\{0,1\}^{n\times n}$ }

\State{$\qquad$and $\mathbf{B}_{i}\sim\mathrm{Unif}(\mathbb{F}_{q}^{n-k_{i}})$ using $\mathfrak{G}$}

\EndFor

\State{Decode $\hat{\mathbf{Y}}$ conditional on $(\boldsymbol{\Pi}_{L},\mathbf{B}_{L})$ from $\mathbf{M}$}

\State{\Return{$\hat{\mathbf{Y}}$}}

\end{algorithmic}

\caption{Rejection-sampled syndrome encoder\label{alg:RSSE}}
\end{algorithm}

We now check that the RSSE simulates an AXN channel. \smallskip{}

\begin{prop}
\label{prop:rsse_noise}The RSSE simulates an AXN channel, with noise distribution
\[
\mathbb{P}(\hat{\mathbf{Z}}=\hat{\mathbf{z}})=\sum_{i=1}^{\infty}\mathbb{P}(L\ge i)\mathbb{E}\left[\kappa_{i}(\hat{\mathbf{z}}\boldsymbol{\Pi}_{i}\,|\,\mathbf{S}_{i})\right]
\]
where $\boldsymbol{\Pi}_{i}\in\{0,1\}^{n\times n}$ is a uniformly random permutation matrix, independent of $\mathbf{S}_{i}\sim\mathrm{Unif}(\mathbb{F}_{q}^{n-k_{i}})$, where $L\in\mathbb{N}$ is the index selected by RSSE, with $\mathbb{P}(L\ge i)$ given recursively as
\[
\mathbb{P}(L\ge i+1)=\mathbb{P}(L\ge i)\mathbb{E}\left[\kappa_{i}(\mathrm{e}\,|\,\mathbf{S}_{i})\right].
\]
\end{prop}
\begin{IEEEproof}
At iteration $i$ (i.e., conditional on $L\ge i$), using the same arguments as Proposition \ref{prop:sse_noise}, the encoder selects $\hat{\mathbf{z}}$ with probability
\[
\mathbb{P}(\hat{\mathbf{Z}}=\hat{\mathbf{z}},\,L=i\,|\,L\ge i)=\mathbb{E}\left[\kappa_{i}(\hat{\mathbf{z}}\boldsymbol{\Pi}_{i}\,|\,\mathbf{S}_{i})\right].
\]
Also, the probability of rejection at iteration $i$ is
\[
\mathbb{P}(L\ge i+1\,|\,L\ge i)=\mathbb{E}\left[\kappa_{i}(\mathrm{e}\,|\,\mathbf{S}_{i})\right].
\]
\end{IEEEproof}
\smallskip{}

\subsection{$(\mathbf{H}, \kappa)$-RSSE}

Consider the special RSSE constructed based on a single $(n, k)$-syndrome coder $(\mathbf{H}, \kappa)$, where $\mathbf{H}_i = \mathbf{H}$ for all $i$, $\kappa_1 = \kappa_2 = \cdots$ and $\kappa_i$ satisfies for any $\mathbf{s} \in \mathbb{F}_{q}^{n-k}$ and $\mathbf{z} \in \mathbb{F}_{q}^{n}$, $\kappa_{i}(\mathbf{z}|\mathbf{s}) = \beta_{\mathbf{z},\mathbf{s}} \kappa(\mathbf{z}|\mathbf{s})$ for some $\beta_{\mathbf{z},\mathbf{s}} \in [0,1]$.
We call this the \emph{$(\mathbf{H}, \kappa)$-RSSE or pure RSSE}.
The constraint of $\kappa_i$ implies that $\hat{\mathbf{V}}_{i}\in\mathbb{F}_{q}^{n}\cup\{\mathrm{e}\}$ can be generated by two steps: 
\begin{enumerate}
    \item Generate $\bar{\mathbf{V}}_{i}\in\mathbb{F}_{q}^{n}$ (using $(\mathbf{H}, \kappa)$) following the  distribution $\kappa(\bar{\mathbf{v}}\,|\,\mathbf{S}_i)$, where $\mathbf{S}_i = \mathbf{B}_{i}-\mathbf{X}\boldsymbol{\Pi}_{i}\mathbf{H}_{i}^{\top}$; 
    \item Accept or reject $\bar{\mathbf{V}}_{i}$, i.e., set 
    \begin{align} \label{eq:accept_or_reject}
        \hat{\mathbf{V}}_{i} = 
        \begin{cases}
            \bar{\mathbf{V}}_{i}, &\text{if } U_i \leq \beta_{\bar{\mathbf{V}}_{i}, \mathbf{S}_i};\\
            \mathrm{e}, &\text{otherwise},
        \end{cases}
    \end{align}
    where $U_i \overset{\text{i.i.d.}}{\sim} \mathrm{Unif}(0,1)$.
\end{enumerate}

We now apply the $(\mathbf{H}, \kappa)$-RSSE to simulate an AXN channel with noise distribution $p_\mathbf{Z}$ 
As both $\hat{\mathbf{Z}}$ and $\mathbf{Z}$ are exchangeable, their probability mass functions depend only on their types.
Hence, $\hat{\mathbf{Z}}$ follows the same distribution as $\mathbf{Z}$ if and only if $\mathrm{type}(\hat{\mathbf{Z}})$ follows the same distribution as $\mathrm{type}(\mathbf{Z})$.
Since permuting a sequence does not change its type, we have $\mathrm{type}(\hat{\mathbf{Z}}) = \mathrm{type}(\hat{\mathbf{V}}_L\Pi_L^{-1}) = \mathrm{type}(\hat{\mathbf{V}}_L) = \mathrm{type}(\bar{\mathbf{V}}_L)$.
Let $\mathbf{T} := \mathrm{type}(\mathbf{Z})$.
Then the simulation task reduces to ensuring $\mathrm{type}(\bar{\mathbf{V}}_L)$ follows $p_\mathbf{T}$.
Lemma \ref{thm:ref_distr} characterizes the distribution of $\bar{\mathbf{T}}_i := \mathrm{type}(\bar{\mathbf{V}}_i)$ for $i \in \mathbb{N}$.

\begin{lem} \label{thm:ref_distr}
The types $\bar{\mathbf{T}}_1, \bar{\mathbf{T}}_2, \ldots$ are i.i.d. with distribution $p_{\bar{\mathbf{T}}}$, which is given by
\begin{equation*}
    p_{\bar{\mathbf{T}}}(\mathbf{t}) := q^{k-n} \tbinom{n}{n\mathbf{t}} \mathbb{E}[\kappa(\mathbf{Z}\,|\,\mathbf{Z}\mathbf{H}^{\top})\,|\,\mathbf{T} = \mathbf{t}]
\end{equation*}
for all $\mathbf{t} \in \mathcal{P}_n(\mathbb{F}_q)$.
\end{lem}

\begin{IEEEproof}
Based on the i.i.d. nature of the common randomness $(\boldsymbol{\Pi}_i, \mathbf{B}_i)_{i \in \mathbb{N}}$, $\mathbf{B}_{i}-\mathbf{X}\boldsymbol{\Pi}_{i}\mathbf{H}^{\top}$ for $i \in \mathbb{N}$ are i.i.d.
Further, $\bar{\mathbf{V}}_i \sim \kappa(\bar{\mathbf{v}}\,|\,\mathbf{B}_i-\mathbf{X}\boldsymbol{\Pi}_i\mathbf{H}^{\top})$ for $i \in \mathbb{N}$ are i.i.d. as well as $\bar{\mathbf{T}}_i = \mathrm{type}(\bar{\mathbf{V}}_i)$ for $i \in \mathbb{N}$.

For any $\mathbf{t} \in \mathcal{P}_n(\mathbb{F}_q)$, we have
\begin{align*}
    \mathbb{P}(\bar{\mathbf{T}}_i = \mathbf{t}) &= \sum_{\mathbf{z} \in \mathbb{F}_q^n\{\mathbf{t}\}} \mathbb{P}(\bar{\mathbf{V}}_i = \mathbf{z}) \\ \displaybreak[0]
    &= \sum_{\mathbf{z} \in \mathbb{F}_q^n\{\mathbf{t}\}} \mathbb{E}[\kappa(\mathbf{z}\,|\,\mathbf{B}_{i}-\mathbf{X}\boldsymbol{\Pi}_{i}\mathbf{H}^{\top})] \\ \displaybreak[0]
    &\overset{\text{(a)}}{=} \sum_{\mathbf{z} \in \mathbb{F}_q^n\{\mathbf{t}\}}\sum_{\mathbf{s} \in \mathbb{F}_q^{n-k}} q^{k-n}\kappa(\mathbf{z}\,|\,\mathbf{s}) \\ \displaybreak[0]
    &\overset{\text{(b)}}{=} \sum_{\mathbf{z} \in \mathbb{F}_q^n\{\mathbf{t}\}} q^{k-n}\kappa(\mathbf{z}\,|\,\mathbf{z}\mathbf{H}^{\top}) \\ \displaybreak[0]
    &= \sum_{\mathbf{z} \in \mathbb{F}_q^n} \mathbbm{1}\{\mathrm{type}(\mathbf{z}) = \mathbf{t}\}q^{k-n}\kappa(\mathbf{z}\,|\,\mathbf{z}\mathbf{H}^{\top}) \\ \displaybreak[0]
    &= \sum_{\mathbf{z} \in \mathbb{F}_q^n} \mathbb{P}(\mathbf{Z} = \mathbf{z}\,|\,\mathbf{T} = \mathbf{t})\tbinom{n}{n\mathbf{t}} q^{k-n}\kappa(\mathbf{z}\,|\,\mathbf{z}\mathbf{H}^{\top}) \\ \displaybreak[0]
    &= q^{k-n}\tbinom{n}{n\mathbf{t}} \mathbb{E}[\kappa(\mathbf{Z}\,|\,\mathbf{Z}\mathbf{H}^{\top})\,|\,\mathbf{T} = \mathbf{t}],
\end{align*}
where (a) holds because $\mathbf{B}_{i}-\mathbf{X}\boldsymbol{\Pi}_{i}\mathbf{H}^{\top} \sim \mathrm{Unif}(\mathbb{F}_q^{n-k})$;
and (b) follows from Definition \ref{def:syndrome_coder}.
\end{IEEEproof}
\smallskip{}

Let 
\[
\mu_{\mathbf{t}}(\mathbf{H}, \kappa) := \mathbb{E}[\kappa(\mathbf{Z}\,|\,\mathbf{Z}\mathbf{H}^{\top})\,|\,\mathbf{T} = \mathbf{t}]
\]
be a parameter of the syndrome coder $(\mathbf{H}, \kappa)$.
It represents the probability of correct decoding when $(\mathbf{H}, \kappa)$ is used as a syndrome decoder for the constant type channel defined by $\mathbf{t}$.
For simplicity, we abbreviate $\mu_{\mathbf{t}}(\mathbf{H}, \kappa)$ as $\mu_{\mathbf{t}}$ in this section.

Based on Lemma \ref{thm:ref_distr}, the simulation task is further transformed to generating a sample from $p_{\mathbf{T}}$ using the i.i.d. samples $\bar{\mathbf{T}}_1, \bar{\mathbf{T}}_2, \ldots$ drawn from $p_{\bar{\mathbf{T}}}$, which can be implemented via the rejection sampling 
\cite{devroye1986nonuniform} with target distribution $p_{\mathbf{T}}$ and proposal distribution $p_{\bar{\mathbf{T}}}$.
Using the standard rejection sampling argument \cite[Section 3.2]{li2024channel}, we obtain Corollary \ref{thm:RS_P_Q}.

\begin{cor} \label{thm:RS_P_Q}
Suppose $p_{\mathbf{T}} \ll p_{\bar{\mathbf{T}}}$, i.e., $p_{\mathbf{T}}$ is absolutely continuous with respect to $p_{\bar{\mathbf{T}}}$.
The rejection sampling with target distribution $p_{\mathbf{T}}$ and proposal distribution $p_{\bar{\mathbf{T}}}$ achieves a maximal acceptance probability of $q^{k-n} \tilde{p}_{\mathbf{T}}(\mathbf{t}^*)^{-1} \mu_{\mathbf{t}^*}$, where 
\begin{align*}
    \tilde{p}_{\mathbf{T}}(\mathbf{t}) := \tbinom{n}{n\mathbf{t}}^{-1} p_{\mathbf{T}}(\mathbf{t})
\end{align*}
for all $\mathbf{t} \in \mathcal{P}_n(\mathbb{F}_q^n)$ and 
\begin{equation*}
    \mathbf{t}^* :=  \underset{\mathbf{t} \in \mathrm{supp}(p_{\mathbf{T}})}{\mathrm{\arg \min}} \tilde{p}_{\mathbf{T}}(\mathbf{t})^{-1}\mu_{\mathbf{t}}.
\end{equation*}
Let the number of iterations $L$ be
\begin{equation} \label{eq:accept_rule}
    L := \min \left\{
    i \in \mathbb{N}:\,U_i \leq
    \tfrac{\tilde{p}_{\mathbf{T}}(\bar{\mathbf{T}}_i)}{\tilde{p}_{\mathbf{T}}(\mathbf{t}^*)} 
    \tfrac{\mu_{\mathbf{t}^*}}{\mu_{\bar{\mathbf{T}}_i}}
    \right\},
\end{equation}
where $U_1, U_2, \ldots \overset{\text{i.i.d.}}{\sim} \mathrm{Unif}(0,1)$ independent of $\bar{\mathbf{T}}_1, \bar{\mathbf{T}}_2, \ldots$
Then $L$ follows the geometric distribution with parameter $q^{k-n} \tilde{p}_{\mathbf{T}}(\mathbf{t}^*)^{-1} \mu_{\mathbf{t}^*}$ and satisfies
\begin{equation*}
    \begin{cases}
        \mathbb{E}[L] = q^{n-k} \tilde{p}_{\mathbf{T}}(\mathbf{t}^*) \mu_{\mathbf{t}^*}^{-1},\\
        H(L) \leq (n-k)\log q + \log \tilde{p}_{\mathbf{T}}(\mathbf{t}^*) \mu_{\mathbf{t}^*}^{-1} + \log e.
    \end{cases}
\end{equation*}
\end{cor}

Equation \eqref{eq:accept_rule} defines the acceptance rule for $\bar{\mathbf{V}}_i$.
Hence, for $\mathbf{s} \in \mathbb{F}_{q}^{n-k}$ and $\mathbf{z} \in \mathbb{F}_{q}^{n}$, we have
\begin{equation*}
    \kappa_{i}(\mathbf{z}|\mathbf{s}) = \tfrac{\tilde{p}_{\mathbf{T}}(\mathrm{type}(\mathbf{z}))}{\tilde{p}_{\mathbf{T}}(\mathbf{t}^*)} 
    \tfrac{\mu_{\mathbf{t}^*}}{\mu_{\mathrm{type}(\mathbf{z})}} \kappa(\mathbf{z}|\mathbf{s}).
\end{equation*}
Note that the expected number of iterations $\mathbb{E}[L]$ reflects the average computational complexity, as it corresponds to the average number of invocations of the syndrome coder $(\mathbf{H}, \kappa)$.
Moreover, the expected communication cost is upper-bounded by $k \log q + H(L) + 1$ if $L$ is encoded via an optimal prefix-free code.
In practice, the Golomb code is well-suited for encoding $L$, as it can achieve optimal lossless compression for geometrically distributed variables \cite{gallagerOptimalSourceCodes1975} with efficient encoding and decoding.

The following presents an example of simulating RTCs using the $(\mathbf{H}, \kappa)$-RSSE, where we assume $p_{\mathbf{T}} \ll p_{\bar{\mathbf{T}}}$.

\begin{eg} \label{eg:RTC}
Consider simulating an RTC defined by $\mathcal{T} \subseteq \mathcal{P}_n(\mathbb{F}_q)$, where $p_{\mathbf{T}}(\mathbf{t}) = \mathbbm{1}\{\mathbf{t} \in \mathcal{T}\}\tbinom{n}{n\mathbf{t}}|\mathbb{F}_q^n\{\mathcal{T}\}|^{-1}$.
In this case, the acceptance rule \eqref{eq:accept_rule} can be written as
\begin{equation} \label{eq:accept_rule_RTC}
    L = \min \big\{
    i \in \mathbb{N}:\,\bar{\mathbf{T}}_i \in \mathcal{T} \text{ and } U_i \leq \tfrac{\mu_{\mathbf{t}^*}}{\mu_{\bar{\mathbf{T}}_i}}
    \big\},
\end{equation}
where $\mathbf{t}^* := \arg\min_{\mathbf{t} \in \mathcal{T}} \mu_{\mathbf{t}}$.
The expected number of iterations is
\begin{align*}
    \mathbb{E}[L] = 2^{n \log q - \log |\mathbb{F}_q^n\{\mathcal{T}\}| - k \log q} \mu_{\mathbf{t}^*}^{-1}.
\end{align*}
The expected number of bits of communication for sending $(L, \mathbf{M})$ is upper-bounded by
\begin{equation*}
    n \log q - \log|\mathbb{F}_q^n\{\mathcal{T}\}| - \log \mu_{\mathbf{t}^*} + \log e + 1.
\end{equation*}
Note that $n \log q - \log|\mathbb{F}_q^n\{\mathcal{T}\}| = n \log q - H(\mathbf{Z})$ is the channel capacity, since $\mathbf{Z} \sim \mathrm{Unif}(\mathbb{F}_q^n\{\mathcal{T}\})$.
If $\mu_{\mathbf{t}^*} = 1$, i.e., $\mu_\mathbf{t} = 1$ for all $\mathbf{t} \in \mathcal{T}$, then the expected communication cost is within $\log e + 1$ bits from the capacity. 

If $\mathcal{T} = \{\mathbf{t}\}$, i.e., the channel reduces to a constant type channel, then the acceptance rule \eqref{eq:accept_rule} can be simplified as 
\begin{equation} \label{eq:accept_rule_CTC}
    L = \min \{i \in \mathbb{N}:\, \bar{\mathbf{T}}_i = \mathbf{t}\}.
\end{equation}
\end{eg}

\begin{rem} \label{rem:mu_t}
The acceptance rule \eqref{eq:accept_rule} as well as \eqref{eq:accept_rule_RTC} generally requires knowing the parameters $\mu_{\mathbf{t}}$ for all $\mathbf{t} \in \mathrm{supp}(p_{\mathbf{T}})$. 
Although $\mu_{\mathbf{t}}$ is computable in principle, its computation becomes prohibitive for large $n$.
Consequently, the acceptance rules \eqref{eq:accept_rule} and \eqref{eq:accept_rule_RTC} are theoretically implementable but not always practical.
A notable exception occurs when $(\mathbf{H}, \kappa)$ is a sufficiently strong syndrome coder that ensures $\mu_{\mathbf{t}} = 1$ for all $\mathbf{t} \in \mathrm{supp}(p_{\mathbf{T}})$.
For example, the trivial $(n, 0)$-syndrome coder with $\mathbf{H} = \mathbf{I}_n$ and $\kappa(\mathbf{z}\,|\,\mathbf{z}) = 1$ for all $\mathbf{z} \in \mathbb{F}_q^n$ satisfies this condition.
Furthermore, for the extreme case of simulating a constant type channel, the acceptance rule \eqref{eq:accept_rule_CTC} does not rely on the value of $\mu_{\mathbf{t}}$.
\end{rem}

\section{State-Dependent Rejection-Sampled Syndrome Encoder} \label{sec:state-RSSE}

Example \ref{eg:RTC} demonstrates that the pure RSSE can simulate any RTC with an almost optimal amount of communication.
To extend this result to general AXN channels, we propose the state-dependent RSSE based on the RTC-decomposition.

\begin{defn} \label{def:RTC_decomp}
A random variable $S$ independent of $\mathbf{X}$ is said to induce an \emph{RTC-decomposition} of an AXN channel with noise distribution $p_\mathbf{Z}$ if for any $s \in \mathrm{supp}(p_S)$, $\mathbf{Z}\,|\,\{S = s\} \sim \mathrm{Unif}(\mathbb{F}_q^n\{\mathcal{T}^{(s)}\})$ for some type set $\mathcal{T}^{(s)} \subseteq \mathcal{P}_n(\mathbb{F}_q)$.
\end{defn}

Note that every AXN channel admits RTC-decompositions.
For example, $S$ can be taken as $\mathbf{T}$.
Based on this, we now introduce the state-dependent RSSE.

\begin{defn} \label{def:state-RSSE}
An \emph{$S$-dependent RSSE} for simulating an AXN channel with noise distribution $p_\mathbf{Z}$ consists of:
\begin{enumerate}
    \item A state $S$, included in the common randomness, that induces an RTC-decomposition of the AXN channel;
    \item A set of $(\mathbf{H}^{(s)}, \kappa^{(s)})$-RSSEs for $s \in \mathrm{supp}(p_S)$, each simulating the RTC with noise distribution $p_{\mathbf{Z}|S=s}$.
\end{enumerate}
The simulation operates by first generating $S \sim p_S$ and then executing the $(\mathbf{H}^{(S)}, \kappa^{(S)})$-RSSE.
\end{defn}

Similar to the generation of $(\boldsymbol{\Pi}_i, \mathbf{B}_i)_{i \in \mathbb{N}}$ in the RSSE, $S$ can be generated by two synchronized random number generators at the encoder and the decoder.
It directly follows from the law of total probability that the $S$-dependent RSSE exactly simulates the target AXN channel.
To characterize the scheme's performance, we define a parameter 
\begin{equation*}
    \mu^{(s)} := \min_{\mathbf{t} \in \mathrm{supp}(p_{\mathbf{T}|S=s})} \mu_\mathbf{t}(\mathbf{H}^{(s)}, \kappa^{(s)})
\end{equation*}
for each $(n, k^{(s)})$-syndrome coder $(\mathbf{H}^{(s)}, \kappa^{(s)})$. 

\begin{thm} \label{thm:state-RSSE_bound}
The $S$-dependent RSSE for simulating an AXN channel with noise distribution $p_{\mathbf{Z}}$ gives a number of iterations $L$ satisfying
\begin{equation} \label{eq:state-RSSE_comp}
    \mathbb{E}[L] = \mathbb{E}\big[2^{C^{(S)} - k^{(S)} \log q - \log \mu^{(S)}} \big],
\end{equation}
where $C^{(s)} := n \log q - H(\mathbf{Z}\,|\,S=s)$ is the capacity of the channel with noise distribution $p_{\mathbf{Z}|S=s}$ for $s \in \mathrm{supp}(p_S)$.
The expected number of bits of communication is at most
\begin{equation} \label{eq:state-RSSE_comm}
    C + I(\mathbf{Z};S) + \mathbb{E}[-\log \mu^{(S)}] + \log e + 1,
\end{equation}
where $C := n \log q - H(\mathbf{Z})$ is the channel capacity.
\end{thm}

\begin{IEEEproof}
Conditional on $S = s$, the $S$-dependent RSSE becomes the $(\mathbf{H}^{(s)}, \kappa^{(s)})$-RSSE that simulates the RTC with noise distribution $p_{\mathbf{Z}|S=s}$. 
From Example \ref{eg:RTC}, the conditional expected number of iterations is
\begin{equation*}
    2^{C^{(s)} - k^{(s)} \log q - \log \mu^{(s)}},
\end{equation*}
and the conditional expected number of bits of communication is upper-bounded by
\begin{equation*}
    n \log q - H(\mathbf{Z} \mid S=s) - \log \mu^{(s)} + \log e + 1.
\end{equation*}

Taking expectation over $S$, we obtain the expected number of iterations $\mathbb{E}[L]$ as in \eqref{eq:state-RSSE_comp}.
Similarly, the expected number of bits of communication is upper bounded by 
\begin{align*}
    & n \log q - H(\mathbf{Z}\,|\,S) + \mathbb{E}[-\log \mu^{(S)}] + \log e + 1 \\
    & = n \log q - H(\mathbf{Z}) + I(\mathbf{Z};S) + \mathbb{E}[-\log \mu^{(S)}] + \log e + 1 \\
    & = C + I(\mathbf{Z};S) + \mathbb{E}[-\log \mu^{(S)}] + \log e + 1.
\end{align*}
\end{IEEEproof}
\smallskip{}

\begin{rem} \label{rem:capacity-achieving}
Observing \eqref{eq:state-RSSE_comm}, the third term $\mathbb{E}[-\log \mu^{(S)}]$ can be eliminated by using sufficiently strong syndrome coders to ensure $\mu^{(s)} = 1$ for all $s$ (as discussed in Remark \ref{rem:mu_t}).
Moreover, $I(\mathbf{Z};S) = I(\mathbf{T};S) \leq H(\mathbf{T}) \leq \log(n+1)^q = O(q\log n)$.
Therefore, for any sequence of AXN channels with noise distributions $p_{\mathbf{Z}^{(n)}}$ over $\mathbb{F}_q^n$ and capacities $C^{(n)}$ such that $\tilde{C}:=\lim_{n \to \infty} C^{(n)}/n$ exists, the state-dependent RSSE can achieve the \emph{asymptotic capacity} $\tilde{C}$, in the sense that the communication rate (expected communication cost divided by $n$) can approach $\tilde{C}$.
In particular, for $\mathrm{BSC}(\alpha)$, $H(\mathbf{T}) = H(\mathrm{wt}(\mathbf{Z})) = \frac{1}{2}\log(2\pi e \alpha(1-\alpha)n) + O(\frac{1}{n})$,
which implies the state-dependent RSSE can achieve the optimal redundancy for simulating BSCs \cite{sriramu2024optimal,flamichRedundancyNonSingularChannel2025a}.
\end{rem}

To optimize the expected communication cost, we seek a state $S$ that induces an RTC-decomposition of the target AXN channel and minimizes $I(\mathbf{Z};S) = I(\mathbf{T};S)$.
For this, we construct a specific state $S^*$ as follows.
Let $\mathbf{t}_1, \cdots, \mathbf{t}_m \in \mathcal{P}_n(\mathbb{F}_q)$ ($m := |\mathcal{P}_n(\mathbb{F}_q)|$) be the sequence of types obtained by sorting $\mathcal{P}_n(\mathbb{F}_q)$ in descending order of $\tilde{p}_{\mathbf{T}}(\mathbf{t}) = \tbinom{n}{n\mathbf{t}}^{-1} p_{\mathbf{T}}(\mathbf{t})$, i.e., $\tilde{p}_{\mathbf{T}}(\mathbf{t}_1) \geq \cdots \geq \tilde{p}_{\mathbf{T}}(\mathbf{t}_m)$.
For each $s \in [1:m]$, we define a type set $\mathcal{T}^{(s)} := \{\mathbf{t}_1, \ldots, \mathbf{t}_s\}$ and set $\mathbf{Z}\,|\,\{S^* = s\} \sim \mathrm{Unif}(\mathbb{F}_q^n\{\mathcal{T}^{(s)}\})$.
Accordingly, the probability mass function of $S^*$ is defined as
\begin{equation*} 
    p_{S^*}(s) := (\tilde{p}_{\mathbf{T}}(\mathbf{t}_s) - \tilde{p}_{\mathbf{T}}(\mathbf{t}_{s+1})) |\mathbb{F}_q^n\{\mathcal{T}^{(s)}\}|,
\end{equation*}
where $\tilde{p}_{\mathbf{T}}(\mathbf{t}_{m+1}) := 0$.
The $S^*$-dependent RSSE shares a similar structure to the layered rejection-sampled universal quantizer \cite{ling2024rejection}.
Proposition \ref{prop:S_star} establishes the near-optimality of this construction.

\begin{prop} \label{prop:S_star}
For any AXN channel with noise distribution $p_{\mathbf{Z}}$ and capacity $C$, the excess functional information lower bound (under $\mathbf{X} \sim \mathrm{Unif}(\mathbb{F}_q^n)$) equals $C + I(\mathbf{Z};S^*)$.
Consequently, the $S^*$-dependent RSSE can achieve an expected communication cost within $\log e + 1$ bits from this lower bound.
\end{prop}

\begin{IEEEproof}
Let $\rho\big(x) := -x \log x$.
The excess functional information lower bound is evaluated as
\begin{align*}
    & \quad\,\sum_{\mathbf{y} \in \mathbb{F}_q^n} \int_{0}^{1} \rho\big(\mathbb{P}(p_{\mathbf{Y}|\mathbf{X}}(\mathbf{y}|\mathbf{X}) \geq u)\big) \, du\\ \displaybreak[0]
    &= \sum_{\mathbf{y} \in \mathbb{F}_q^n} \int_{0}^{1} 
    \rho\big(\mathbb{P}(p_{\mathbf{Z}}(\mathbf{y} - \mathbf{X}) \geq u)\big) \, du \\ \displaybreak[0]
    &\overset{(a)}{=} q^n \int_{0}^{1} \rho\big(\mathbb{P}(p_{\mathbf{Z}}(\mathbf{X}) \geq u)\big) \, du \\ \displaybreak[0]
    &= q^n \int_{0}^{1} \rho\big(\mathbb{P}(\tilde{p}_{\mathbf{T}}(\mathrm{type}(\mathbf{X})) \geq u)\big) \, du \\ \displaybreak[0]
    &= q^n \sum_{s=1}^{m} \int_{\tilde{p}_{\mathbf{T}}(\mathbf{t}_{s+1})}^{\tilde{p}_{\mathbf{T}}(\mathbf{t}_{s})} \rho\big(\mathbb{P}(\tilde{p}_{\mathbf{T}}(\mathrm{type}(\mathbf{X})) \geq u)\big) \, du\\ \displaybreak[0]
    &= q^n \sum_{s=1}^{m} \int_{\tilde{p}_{\mathbf{T}}(\mathbf{t}_{s+1})}^{\tilde{p}_{\mathbf{T}}(\mathbf{t}_{s})} \rho\big(\mathbb{P}(\mathrm{type}(\mathbf{X}) \in \mathcal{T}^{(s)})\big) \, du\\ \displaybreak[0]
    &= q^n \sum_{s=1}^{m} (\tilde{p}_{\mathbf{T}}(\mathbf{t}_{s}) - \tilde{p}_{\mathbf{T}}(\mathbf{t}_{s+1})) \rho\big(q^{-n}|\mathbb{F}_q^n\{\mathcal{T}^{(s)}\}|\big) \\ \displaybreak[0]
    &= - \sum_{s=1}^{m} (\tilde{p}_{\mathbf{T}}(\mathbf{t}_{s}) - \tilde{p}_{\mathbf{T}}(\mathbf{t}_{s+1})) |\mathbb{F}_q^n\{\mathcal{T}^{(s)}\}| \log q^{-n}|\mathbb{F}_q^n\{\mathcal{T}^{(s)}\}| \\ \displaybreak[0]
    &= \sum_{s=1}^{m} P_{S^*}(s) \big(n\log q - \log |\mathbb{F}_q^n\{\mathcal{T}^{(s)}\}|\big) \\ \displaybreak[0]
    &= \sum_{s=1}^{m} P_{S^*}(s) (n\log q - H(\mathbf{Z}\,|\,S^* = s)) \\ \displaybreak[0]
    &= n\log q - H(\mathbf{Z}\,|\,S^*) \\ \displaybreak[0]
    &= C + I(\mathbf{Z};S^*),
\end{align*}
where $(a)$ is because $\mathbf{X} \sim \mathrm{Unif}(\mathbb{F}_q^n)$.
\end{IEEEproof}
\bigskip{}


We now show applications of the state-dependent RSSE in the simulation of binary AXN channels.
In such settings, the Hamming weight $W := \mathrm{wt}(\mathbf{Z})$ of the target noise $\mathbf{Z}$ can replace the role of the type $\mathbf{T} := \mathrm{type}(\mathbf{Z})$.
The design task is to define a state $S$ that induces an RTC-decomposition of the target AXN channel, and configure the syndrome coders $(\mathbf{H}^{(s)}, \kappa^{(s)})$ for all $s \in \mathrm{supp}(p_S)$.
We use BCH codes and polar codes to develop practical schemes for the short blocklength and the asymptotic regimes, respectively.

\subsection{Short Blocklength Regime: $S^*$-dependent BCH-RSSE}

As discussed in Remarks \ref{rem:mu_t} and \ref{rem:capacity-achieving}, constructing an $S$-dependent RSSE based on syndrome coders with $\mu^{(s)} = 1$ for all $s \in \mathrm{supp}(p_S)$ not only makes the acceptance rule \eqref{eq:accept_rule_RTC} practical but also eliminates the penalty term $\mathbb{E}[-\log \mu^{(S)}]$ in \eqref{eq:state-RSSE_comm}.
BCH codes \cite{boseClassErrorCorrecting1960, hocquenghemCodesCorrecteursDerreurs1959} have good parameters (i.e., large dimension $k$ and minimum (or designed) distance $d$\,) and efficient decoding algorithms that correct up to $\lfloor (d-1)/2 \rfloor$ errors, such as the Berlekamp-Massey (BM) algorithm \cite{berlekampDecodingBinaryBoseChaudhuriHocquenghem1965, masseyShiftregisterSynthesisBCH1969}.
In this implementation, we employ binary primitive BCH codes together with the BM algorithm to form the syndrome coders $(\mathbf{H}^{(s)}, \kappa^{(s)})$ for $s \in \mathrm{supp}(p_S)$ such that $\mu^{(s)} = 1$.

Let $w^{(s)} := \max \mathrm{supp}(p_{W|S=s})$ denote the maximal Hamming weight of the noise $\mathbf{Z}\,|\,\{S = s\}$, and $d^{(s)}$ denote the designed distance of the BCH code defined by $\mathbf{H}^{(s)}$.
If $d^{(s)} \geq 2w^{(s)}+1$, then $\mu^{(s)} = 1$ is guaranteed.
In addition, according to \eqref{eq:state-RSSE_comp}, to optimize the average complexity, the code dimension $k^{(s)}$ should be as large as possible.
Therefore, for each $s$, we select the BCH code with the largest dimension among all those having a designed distance of at least $2w^{(s)}+1$.
We refer to this scheme as the $S$-dependent BCH-RSSE.

\begin{figure*}[htbp!]
    \begin{centering}

        \includegraphics[width=0.75\linewidth]{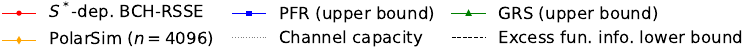}

        \begin{subfigure}{0.6\linewidth}
            \centering
            \includegraphics[width=\textwidth]{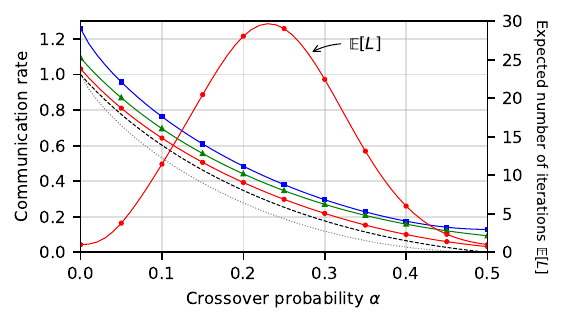}
            
            \vspace{-2ex}
            \caption{$n=31$}
        \end{subfigure}

        \begin{subfigure}{0.6\linewidth}
            \centering
            \includegraphics[width=\textwidth]{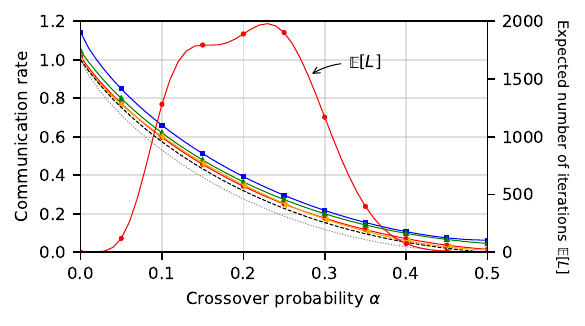}

            \vspace{-2ex}
            \caption{$n=63$}
            \label{fig:BCH-RSSE_n63}
        \end{subfigure}

    \end{centering}
    \caption{The communication rate and expected number of iterations given by the $S^*$-dependent BCH-RSSE for simulating $n$ copies of $\mathrm{BSC}(\alpha)$ with $n=31, 63$ and $\alpha\in[0,\frac{1}{2}]$.}
    \label{fig:BCH-RSSE}
    \vspace{-3ex}
\end{figure*}

Figure \ref{fig:BCH-RSSE} shows the communication rate\footnote{The geometrically distributed variable $L\,|\,\{S^* = s\}$ is encoded using an optimal prefix-free code. The expected encoding length is given by \cite[equation (8)]{gallagerOptimalSourceCodes1975}, where the ceiling $\lceil \cdot \rceil$ should be corrected to a floor $\lfloor \cdot \rfloor$.} and the expected number of iterations for simulating $n$ copies of $\mathrm{BSC}(\alpha)$ via the $S^*$-dependent BCH-RSSE, where $n = 31$ and $63$, $\alpha\in[0,\frac{1}{2}]$.
For comparison, we also plot communication rates of the following schemes:
\begin{enumerate}
\item PFR \cite{sfrl_trans}: Only the upper bound \eqref{eq:PFR_bound} is plotted due to the high computational cost at the given blocklength $n$;
\item GRS \cite{harsha2010communication}: Only the upper bound \cite[Theorem V.2]{goc2024causal} is plotted for the same reason as above;
\item PolarSim \cite{sriramu2024fast}: Experiment results for $n = 4096$ are plotted in Figure \ref{fig:BCH-RSSE_n63};
\item Lower bounds: Channel capacity and the excess functional information lower bound \cite[Proposition 1]{sfrl_trans}.
\end{enumerate}

As seen from Figure \ref{fig:BCH-RSSE}, the $S^*$-dependent BCH-RSSE outperforms both PFR and GRS.
It achieves a rate close to the excess functional information lower bound, with a gap less than $0.045$ for $n = 31$ and $0.024$ for $n = 63$ across all $\alpha$.
Moreover, the $S^*$-dependent BCH-RSSE with $n = 63$ even performs comparably to the PolarSim with $n = 4096$, while its much smaller blocklength enables lower communication latency.
Regarding the computational complexity, the expected number of iterations remains below $30$ for $n = 31$ and $2000$ for $n = 63$ across all $\alpha$, which is practical owing to the efficiency of the BM algorithm.
It should be pointed out that for specific BCH codes such as Hamming codes, simplex codes and repetition codes, decoding can be simpler than the BM algorithm.
Hence, the actual average complexity can be reduced.

\subsection{Asymptotic Regime: $W$-dependent Polar-RSSE} \label{sec:Polar-RSSE}

Although the $S^*$-dependent BCH-RSSE achieves an almost optimal communication cost, its computational complexity grows exponentially in $n$, limiting its application in the asymptotic regime.
Polar codes are well-known for achieving the capacity for binary-input memoryless symmetric channels under successive cancellation (SC) decoding \cite{arikan2009channel}, and the rate-distortion function for binary symmetric sources under SC encoding \cite{koradaPolarCodesAre2010}, both with $O(n \log n)$ complexity.
To address the limitation of the state-dependent BCH-RSSE, we employ polar codes for lossy source coding to construct a scheme.

Fix a target distortion $D \in (0, 1/2)$.
Let $(\mathbf{H}^{(n)}, \kappa^{(n)})$ be the randomized SC encoder \cite{koradaPolarCodesAre2010} of the polar code with length $n$ and rate $R^{(n)}$ designed for lossy source coding of a binary symmetric source with average distortion $D$.
Note that $n$ is restricted to powers of two throughout this subsection.
The $(\mathbf{H}^{(n)}, \kappa^{(n)})$-RSSE, also called the $(n, R^{(n)})$-polar-RSSE, is limited to the acceptance rule \eqref{eq:accept_rule_CTC} because the parameter $\mu_{\mathbf{t}}(\mathbf{H}^{(n)}, \kappa^{(n)})$ is generally difficult to compute (see Remark \ref{rem:mu_t}).
As a result, it is only applicable to simulating a constant weight channel $\mathrm{CWC}^{(n)}(w)$, where the noise is uniform over $\mathbb{F}_2^n\{w\}$ for a fixed $w \in [0:n]$.
Using results from \cite{koradaPolarCodesAre2010}, we show that the polar-RSSE achieves the asymptotic capacity for a particular sequence of constant weight channels with polynomial complexity. The proof is in Appendix~\ref{pf:thm_polar_RSSE}.

\begin{thm} \label{thm:polar-RSSE}
Fix $D \in [0, 1/2]$.
For any sequence $(w^{(n)})_n$ with $w^{(n)} \!\in [0:n]$ satisfying $|w^{(n)} - nD| = O(\sqrt{n})$ and any $R > 1 - H_\mathrm{b}(D)$, 
there exists a sequence of $(n, R^{(n)})$-polar-RSSEs with $R^{(n)} < R$ that simulate constant weight channels $\mathrm{CWC}^{(n)}(w^{(n)})$.
They achieve a communication rate of $R^{(n)} + O(\log n/n)$ with expected $O(\sqrt{n})$ SC encoding iterations.
\end{thm}

The asymptotic capacity of the constant weight channels in Theorem \ref{thm:polar-RSSE} is $\lim_{n \to \infty} \left(n - \log\tbinom{n}{w^{(n)}}\right)\!/n = 1 - H_\mathrm{b}(D)$.
Furthermore, based on $\hat{\mathbf{Y}} + \mathbf{1} = \mathbf{X} + \hat{\mathbf{Z}} + \mathbf{1}$ and $\mathrm{wt}(\hat{\mathbf{Z}} + \mathbf{1}) = n - \mathrm{wt}(\hat{\mathbf{Z}})$, $\mathrm{CWC}^{(n)}(w)$ can be simulated using a scheme designed for $\mathrm{CWC}^{(n)}(n-w)$ by flipping all output bits.
This allows us to extend the domain of $D$ in Theorem \ref{thm:polar-RSSE} to $[0, 1]$.

We now construct the $W^{(n)}$-dependent polar-RSSE for simulating $n$ copies of $\mathrm{BSC}(\alpha)$, where $W^{(n)}$ is the Hamming weight of the noise and follows the binomial distribution with parameters $n$ and $\alpha$.
The design task is reduced to configuring $(\mathbf{H}^{(n,w)}, \kappa^{(n,w)})$ for each $w \in [0:n]$, which defines a pure RSSE for simulating $\mathrm{CWC}^{(n)}(w)$.
Fix any $\epsilon > 0$.
Define
\begin{equation*}
    \mathcal{W}_\epsilon^{(n)} := \big\{w \in [0:n]:\, |w - n\alpha| \geq \sqrt{\alpha(1-\alpha)n/\epsilon}\big\}.
\end{equation*}
By Chebyshev's inequality, $\mathbb{P}(W^{(n)} \!\in \mathcal{W}_\epsilon^{(n)}) \leq \epsilon$.
For each $w$, let $(\mathbf{H}^{(n,w)}, \kappa^{(n,w)})$ be:
\begin{itemize}
    \item The complete $(n, n)$-syndrome coder, i.e., $\mathbf{H}^{(n,w)} = \mathbf{0}$ and $\kappa^{(n,w)}: \{0\} \to \mathbb{F}_2^n\{w\}$, if $w \in \mathcal{W}_\epsilon^{(n)}$;
    \item The randomized SC encoder of the $(n, R^{(n)})$-polar code designed for target distortion $\alpha$, otherwise.
\end{itemize}
Based on Theorem \ref{thm:polar-RSSE}, there exists a sequence of polar codes with $R^{(n)} < 1 - H_\mathrm{b}(\alpha) + \epsilon$ such that the construction achieves a communication rate of
\begin{align*}
    & \mathbb{P}(W^{(n)} \!\in \mathcal{W}_\epsilon^{(n)}) + \mathbb{P}(W^{(n)} \!\notin \mathcal{W}_\epsilon^{(n)}) (R^{(n)} + O(\log n/n))\\ 
    & \leq \, \epsilon + R^{(n)} + O(\log n/n),
\end{align*}
with expected $\mathbb{P}(W^{(n)} \!\notin \mathcal{W}_\epsilon^{(n)})\, O(\sqrt{n}) = O(\sqrt{n})$ SC encoding iterations.
This shows the $W^{(n)}$-dependent polar-RSSE achieves the asymptotic capacity for simulating BSCs with polynomial complexity.

\section{Greedy Rejection-Sampled Syndrome Encoder\label{sec:grsse}}

In this section, we will give an algorithm for the construction of an RSSE for the exact simulation of an AXN channel with noise distribution $p_{\mathbf{Z}}(\mathbf{z})$, called \emph{greedy rejection-sampled syndrome encoder (GRSSE)}. It is inspired by the greedy rejection sampling scheme \cite{harsha2010communication}, which maximizes the probability of acceptance at iteration $1$, then maximizes the probability of acceptance at iteration $2$, and so on. Unlike greedy rejection sampling, we do not have a fixed proposal distribution, but rather has the flexibility to design $\kappa_{i}$ which helps us further maximize the probability of acceptance.

\medskip{}

\begin{defn}
Given a sequence of parity-check matrices $(\mathbf{H}_{i})_{i}$ and the target exchangeable noise distribution $p_{\mathbf{Z}}$, the \emph{greedy rejection-sampled syndrome encoder (GRSSE)} $(\mathbf{H}_{i},\kappa_{i})_{i}$ is defined recursively as $p_{\mathbf{Z}}^{(1)}(\mathbf{z})=p_{\mathbf{Z}}(\mathbf{z})$ for $\mathbf{z}\in\mathbb{F}_{q}^{n}$, and for $i=1,2,\ldots$, using the notations in Proposition \ref{prop:rsse_noise},
\begin{equation}
\kappa_{i}=\underset{\kappa'_{i}:\,\mathbb{E}[\kappa'_{i}(\mathbf{z}\boldsymbol{\Pi}_{i}\,|\,\mathbf{S}_{i})]\le p_{\mathbf{Z}}^{(i)}(\mathbf{z}),\,\forall\mathbf{z}\in\mathbb{F}_{q}^{n}}{\mathrm{argmin}}\mathbb{E}[\kappa'_{i}(\mathrm{e}|\mathbf{S}_{i})],\label{eq:grsse_argmin}
\end{equation}
where the argmin is over $\kappa'_{i}$ satisfying Definition \ref{def:rsse} (if the argmin is not unique, select $\kappa_{i}$ arbitrarily), and 
\[
p_{\mathbf{Z}}^{(i+1)}(\mathbf{z})=\frac{p_{\mathbf{Z}}^{(i)}(\mathbf{z})-\mathbb{E}[\kappa_{i}(\mathbf{z}\boldsymbol{\Pi}_{i}\,|\,\mathbf{S}_{i})]}{\mathbb{E}[\kappa_{i}(\mathrm{e}\,|\,\mathbf{S}_{i})]}.
\]
We can terminate this process when $\mathbb{E}[\kappa_{i}(\mathrm{e}\,|\,\mathbf{S}_{i})]=0$. 
\end{defn}
\smallskip{}

Intuitively, GRSSE first selects $\kappa_{1}$ that minimizes $\mathbb{E}[\kappa_{1}(\mathrm{e}\,|\,\mathbf{S}_{1})]$, the probability of rejection at iteration $1$ (see Proposition \ref{prop:rsse_noise}). Note that we require $\mathbb{E}[\kappa_{1}(\mathbf{z}\boldsymbol{\Pi}_{1}\,|\,\mathbf{S}_{1})]=\mathbb{P}(\hat{\mathbf{Z}}=\mathbf{z},\,L=1)\le p_{\mathbf{Z}}(\mathbf{z})$ if we want to exactly simulate $p_{\mathbf{Z}}$, or else the probability of the noise $\mathbf{z}$ will be too large. Hence, we minimize $\mathbb{E}[\kappa_{1}(\mathrm{e}\,|\,\mathbf{S}_{1})]$ subject to $\mathbb{E}[\kappa_{1}(\mathbf{z}\boldsymbol{\Pi}_{1}\,|\,\mathbf{S}_{1})]\le p_{\mathbf{Z}}(\mathbf{z})$ for all $\mathbf{z}$. The gaps $p_{\mathbf{Z}}(\mathbf{z})-\mathbb{E}[\kappa_{1}(\mathbf{z}\boldsymbol{\Pi}_{1}\,|\,\mathbf{S}_{1})]$, normalized to be a probability distribution $p_{\mathbf{Z}}^{(2)}(\mathbf{z})=p_{\mathbf{Z}|L\ge2}(\mathbf{z})$, must be filled in the remaining iterations to ensure exact simulation. We then continue this process for iteration $2$ to simulate the noise distribution $p_{\mathbf{Z}}^{(2)}(\mathbf{z})$, and so on.
It can be checked that $p_{\mathbf{Z}}^{(i)}(\mathbf{z})=p_{\mathbf{Z}|L\ge i}(\mathbf{z})$.

We now describe the computation of the argmin in (\ref{eq:grsse_argmin}). There are various symmetries in (\ref{eq:grsse_argmin}) that allow us to reduce the dimension of the problem. First, due to the random permutation $\boldsymbol{\Pi}_{i}$, the distributions $p_{\mathbf{Z}}^{(i)}(\mathbf{z})$  are always exchangeable, and hence we only need to keep track of the distribution of the type $\mathbf{t}=\mathrm{type}(\mathbf{z})$, i.e., we consider $p_{\mathbf{T}}^{(i)}(\mathbf{t})=\sum_{\mathbf{z}:\mathrm{type}(\mathbf{z})=\mathbf{t}}p_{\mathbf{Z}}^{(i)}(\mathbf{z})$ for the distribution of $\mathbf{T}=\mathrm{type}(\mathbf{Z})\in\mathcal{P}_{n}(\mathbb{F}_{q})$ when $\mathbf{Z}\sim p_{\mathbf{Z}}^{(i)}$, and consider a conditional distribution $\tilde{\kappa}_{i}(\mathbf{t}|\mathbf{s})$ from the syndrome to the type instead of $\kappa_{i}(\mathbf{z}|\mathbf{s})$. This significantly reduces the dimension of the problem since $|\mathcal{P}_{n}(\mathbb{F}_{q})|=O(n^{q-1})\ll|\mathbb{F}_{q}^{n}|=q^{n}$. Second, the constraint in Definition \ref{def:rsse} requires $\tilde{\kappa}_{i}(\mathbf{t}|\mathbf{s})>0$ only if there exists $\mathbf{z}$ with $\mathrm{type}(\mathbf{z})=\mathbf{t}$ and $\mathbf{z}\mathbf{H}_{i}^{\top}=\mathbf{s}$. Define the \emph{type set} of the syndrome $\mathbf{s}$ to be
\[
\mathcal{T}_{\mathbf{H}}(\mathbf{s})=\big\{\mathrm{type}(\mathbf{z}):\,\mathbf{z}\in\mathbb{F}_{q}^{n},\,\mathbf{z}\mathbf{H}^{\top}=\mathbf{s}\big\}\subseteq\mathcal{P}_{n}(\mathbb{F}_{q}).
\]
Then syndromes with the same type set are subject to the same constraint, and hence can be considered together. The number of different type sets is often significantly smaller than the number of syndromes. For example, the $[24,12,8]$ Golay code \cite{golay1949digitalcoding} has $2^{12}$ syndromes but only $5$ type sets. Define the \emph{type set distribution} to be the probability mass function 
\begin{equation}
p_{\mathcal{T}_{\mathbf{H}}}(\mathcal{T}):=q^{-(n-k)}|\{\mathbf{s}:\,\mathcal{T}_{\mathbf{H}}(\mathbf{s})=\mathcal{T}\}| \label{eq:typeset_dist}
\end{equation}
over $\mathcal{T}\in \mathcal{T}_{\mathbf{H}}(\mathbb{F}_{q}^{n-k})$, i.e., it is the distribution of $\mathcal{T}_{\mathbf{H}}(\mathbf{S})$ where $\mathbf{S}\sim\mathrm{Unif}(\mathbb{F}_{q}^{n-k})$. For $\mathcal{T}\in \mathcal{T}_{\mathbf{H}_{i}}(\mathbb{F}_{q}^{n-k_i})$, $\mathbf{t}\in\mathcal{P}_{n}(\mathbb{F}_{q})$, let
\[
\gamma_{i,\mathcal{T},\mathbf{t}}:=q^{-(n-k)}\sum_{\mathbf{s},\mathbf{z}:\mathcal{T}_{\mathbf{H}}(\mathbf{s})=\mathcal{T},\,\mathrm{type}(\mathbf{z})=\mathbf{t}}\kappa_{i}(\mathbf{z}|\mathbf{s}).
\]
Note that $\sum_{\mathcal{T},\mathbf{t}}\gamma_{i,\mathcal{T},\mathbf{t}}$ is generally not $1$ since the random mapping $\kappa_{i}$ may have a positive probability of mapping $\mathbf{s}$ to the rejection symbol $\mathrm{e}$. We can maximize the probability of acceptance by maximizing $\sum_{\mathcal{T},\mathbf{t}}\gamma_{i,\mathcal{T},\mathbf{t}}$.
Hence, 
the problem (\ref{eq:grsse_argmin}) is equivalent to the problem of maximizing $\sum_{\mathcal{T},\mathbf{t}}\gamma_{i,\mathcal{T},\mathbf{t}}$ subject to the linear constraints
\[
\begin{array}{ll}
\gamma_{i,\mathcal{T},\mathbf{t}}\ge0,\;\forall \mathcal{T},\mathbf{t};\; & \sum_{\mathbf{t}}\gamma_{i,\mathcal{T},\mathbf{t}}\le p_{\mathcal{T}_{\mathbf{H}_{i}}}(\mathcal{T}),\;\forall \mathcal{T};\\
\sum_\mathcal{T}\gamma_{i,\mathcal{T},\mathbf{t}}\le p_{\mathbf{T}}^{(i)}(\mathbf{t}),\;\forall\mathbf{t};\; & \gamma_{i,\mathcal{T},\mathbf{t}}=0\;\;\text{if}\;\mathbf{t}\notin \mathcal{T}.
\end{array}
\]
This maximization can be solved efficiently using a linear programming or maximum flow algorithm. Refer to Algorithm \ref{alg:GRSSE}, which is for the precomputation of $\kappa_{i}$ in GRSSE. In practice, before the communication commences, the encoder first precomputes the $\kappa_{i}$'s using Algorithm \ref{alg:GRSSE}. During the communication, the encoder and decoder use Algorithm \ref{alg:RSSE} to perform encoding and decoding. Although  Algorithm \ref{alg:GRSSE} may take a significant amount of time, it is only computed once, and the $\kappa_{i}$'s can be reused for subsequent communications. Algorithm \ref{alg:GRSSE} also computes the distribution $p_{L}$, which is useful if we are going to encode $L$ via Huffman coding.

\begin{algorithm}
\textbf{Procedure} $\textsc{GRSSE\_Precompute}(p_{\mathbf{T}}(\mathbf{t}),(\mathbf{H}_{i})_{i}):$

\textbf{$\;\;\;\;$Input:} type distribution $p_{\mathbf{T}}(\mathbf{t})=\sum_{\mathbf{z}:\mathrm{type}(\mathbf{z})=\mathbf{t}}p_{\mathbf{Z}}(\mathbf{z})$,

\textbf{$\;\;\;\;$$\;\;\;\;$$\;\;\;\;$}parity-check matrices $(\mathbf{H}_{i})_{i}$

\textbf{$\;\;\;\;$Output:} RSSE $(\mathbf{H}_{i},\kappa_{i})_{i}$, distribution $p_{L}$

\smallskip{}

\begin{algorithmic}[1]

\State{$p_{\mathbf{T}}^{(1)}(\mathbf{t})\leftarrow p_{\mathbf{T}}(\mathbf{t})$}

\State{$S\leftarrow1$}

\For{$i=1,2,\ldots$}

\State{$p_{\mathcal{T}_{\mathbf{H}_{i}}}(\mathcal{T})\leftarrow q^{-(n-k_i)}|\{\mathbf{s}:\,\mathcal{T}_{\mathbf{H}_i}(\mathbf{s})=\mathcal{T}\}|$, }

\State{$\quad$$\forall \mathcal{T}\in \mathcal{T}_{\mathbf{H}_{i}}(\mathbb{F}_{q}^{n-k_i})$}

\State{Construct weighted directed graph $G$ with}

\State{$\quad$vertices $V=\{\mathrm{src},\mathrm{des}\}\cup \mathcal{T}_{\mathbf{H}_{i}}(\mathbb{F}_{q}^{n-k_i})\cup\mathcal{P}_{n}(\mathbb{F}_{q})$,}

\State{$\quad$edges $(\mathcal{T},\mathbf{t})$ with capacity $\infty$ for $\mathcal{T},\mathbf{t}$ with $\mathbf{t}\in \mathcal{T}$,}

\State{$\quad$edges $(\mathrm{src},\mathcal{T})$ with capacity $p_{\mathcal{T}_{\mathbf{H}_{i}}}(\mathcal{T})$ }

\State{$\qquad$for $\mathcal{T}\in \mathcal{T}_{\mathbf{H}_{i}}(\mathbb{F}_{q}^{n-k_i})$,}

\State{$\quad$edges $(\mathbf{t},\mathrm{des})$ with capacity $p_{\mathbf{T}}^{(i)}(\mathbf{t})$ for $\mathbf{t}\in\mathcal{P}_{n}(\mathbb{F}_{q})$}

\State{Compute maximum flow of $G$ from $\mathrm{src}$ to $\mathrm{des}$}

\State{$F\leftarrow$ total flow, $\;\gamma_{i,\mathcal{T},\mathbf{t}}\leftarrow$ flow along edge $(\mathcal{T},\mathbf{t})$}

\State{$\kappa_{i}(\mathbf{z}|\mathbf{s})\leftarrow0$, $\forall\mathbf{s},\mathbf{z}$}

\State{$\mathrm{vis}(\mathbf{s},\mathbf{t})\leftarrow0$, $\forall\mathbf{s},\mathbf{t}$}

\For{$\mathbf{z}\in\mathbb{F}_{q}^{n}$}

\State{$\mathbf{s}\leftarrow\mathbf{z}\mathbf{H}_{i}^{\top}$, $\;\mathcal{T}\leftarrow \mathcal{T}_{\mathbf{H}_{i}}(\mathbf{s}),$ $\;\mathbf{t}\leftarrow\mathrm{type}(\mathbf{z})$}

\If{$\mathrm{vis}(\mathbf{s},\mathbf{t})=0$}

\State{$\kappa_{i}(\mathbf{z}|\mathbf{s})\leftarrow\gamma_{i,\mathcal{T},\mathbf{t}}/p_{\mathcal{T}_{\mathbf{H}_{i}}}(\mathcal{T})$, $\;\mathrm{vis}(\mathbf{s},\mathbf{t})\leftarrow1$}

\EndIf

\EndFor

\State{$\kappa_{i}(\mathrm{e}|\mathbf{s})\leftarrow1-\sum_{\mathbf{t}\in \mathcal{T}}\gamma_{i,\mathcal{T},\mathbf{t}}/p_{\mathcal{T}_{\mathbf{H}_{i}}}(\mathcal{T})$}

\State{$\quad$(where $\mathcal{T}=\mathcal{T}_{\mathbf{H}_{i}}(\mathbf{s})$)}

\State{$p_{L}(i)\leftarrow SF$}

\If{$F=1$}

\State{\Return{$(\mathbf{H}_{j},\kappa_{j})_{j\in[i]}$, $p_{L}$}}

\EndIf

\State{$p_{\mathbf{T}}^{(i+1)}(\mathbf{t})\leftarrow(p_{\mathbf{T}}^{(i)}(\mathbf{t})-\sum_{\mathcal{T}}\gamma_{i,\mathcal{T},\mathbf{t}})/(1-F)$}

\State{$S\leftarrow S(1-F)$}

\EndFor

\end{algorithmic}

\caption{Precomputation for greedy rejection-sampled syndrome encoder\label{alg:GRSSE}}
\end{algorithm}

\smallskip{}


\section{Theoretical Analysis of GRSSE\label{sec:grsse-1}}

We now analyze the performance of GRSSE. For simplicity, we consider the inner codes $\mathbf{H}_{i}=\mathbf{H}$ to be the same for all $i$. We call this the $\mathbf{H}$\emph{-GRSSE} or \emph{pure GRSSE}. Let the distance of the code be $d:=\min_{\mathbf{z}\neq0:\,\mathbf{z}\mathbf{H}^{\top}=0}\mathrm{wt}(\mathbf{z})$. Theorem \ref{thm:GRSSE_bound} shows that $\mathbf{H}$-GRSSE achieves the capacity under very mild conditions. 
Note that the bound in Theorem \ref{thm:GRSSE_bound} holds regardless of the distribution of the input $\mathbf{X}$.
A proof sketch will be presented later in this section. The complete proof is in Appendix \ref{subsec:pf_GRSSE_bound}.\medskip{}

\begin{thm}
\label{thm:GRSSE_bound}The $\mathbf{H}$-GRSSE for the simulation of the AXN channel with noise distribution $p_{\mathbf{Z}}$, with $\mathbf{H}_{i}=\mathbf{H}$ where the code $\mathbf{H}\in\mathbb{F}_{q}^{(n-k)\times n}$ has a distance $d$, gives a number of iteration $L$ satisfying
\begin{align}
\mathbb{E}[\log L] & \le C-\mathbb{P}\left(\mathrm{wt}(\mathbf{Z})<d/2\right)k\log q+\eta+1,\label{eq:grsse_logl}
\end{align}
where $C=I(\mathbf{X};\mathbf{Y})$ is the channel capacity (where $\mathbf{X}\sim\mathrm{Unif}(\mathbb{F}_{q}^{n})$, $\mathbf{Z}\sim p_{\mathbf{Z}}$, $\mathbf{Y}=\mathbf{X}+\mathbf{Z}$) and $\eta:=(1+e^{-1})\log e$. Hence, the expected number of bits of communication (for sending $L,\mathbf{M}$) is at most
\begin{align}
 & C+\mathbb{P}\left(\mathrm{wt}(\mathbf{Z})\ge d/2\right)k\log q\nonumber \\
 & \;+\log\left(C-\mathbb{P}\left(\mathrm{wt}(\mathbf{Z})<d/2\right)k\log q+\eta+2\right)+\eta+3.\label{eq:grsse_comm}
\end{align}
In particular, if $\mathrm{wt}(\mathbf{Z})<d/2$ almost surely (e.g., the Hamming ball channel and constant weight channel in Section \ref{sec:setting}), the communication is upper-bounded by
\[
C+\log\left(C-k\log q+\eta+2\right)+\eta+3.
\]
\end{thm}
\smallskip{}

To achieve the capacity $C$, we need $d$ to be large enough so $\mathrm{wt}(\mathbf{Z})<d/2$ with high probability. A direct corollary is that for every sequence of AXN channels with noise distributions $p_{\mathbf{Z}^{(n)}}$ over $\mathbb{F}_{q}^{n}$ and capacities $C^{(n)}$ for $n=1,2,\ldots$, with asymptotic capacity $\tilde{C}=\lim_{n\to\infty}C^{(n)}/n$, as long as we find a sequence of inner codes $\mathbf{H}^{(n)}$ with distance $d^{(n)}$ such that $\mathbb{P}(\mathrm{wt}(\mathbf{Z}^{(n)})\ge d^{(n)}/2)\to0$, then GRSSE achieves the asymptotic capacity, in the sense that the expected number of bits of communication per dimension $n$ approaches $\tilde{C}$. As a result, by simulating the sequence of Hamming ball channels $p_{\mathbf{Z}^{(n)}}=\mathrm{Unif}(\{\mathbf{z}\in\mathbb{F}_{q}^{n}:\,\mathrm{wt}(\mathbf{z})\le\alpha n\})$ with $\tilde{C}=\log q+(1-\alpha)\log(1-\alpha)+\alpha\log(\alpha/(q-1))$ for $\alpha \le 1-1/q$, as long as the distances satisfy $\mathrm{liminf}_{n}d^{(n)}/n>2\alpha$, GRSSE can achieve the rate-distortion function for Hamming distortion.

Interestingly, unlike conventional lossy source coding which requires a small covering radius of the code, GRSSE requires a large enough distance $d$. GRSSE does not require $k$ and $d$ to be precise or optimal in any sense, as $d$ only needs to be large enough. The criteria for GRSSE to be capacity-achieving is significantly easier to satisfy compared to channel coding and lossy source coding.

Even the trivial inner code with $k=0$, $\mathbf{H}=\mathbf{I}$ (where $d=\infty$) can achieve the capacity asymptotically, as (\ref{eq:grsse_comm}) gives $C+\log(C+\eta+2)+\eta+3$. In this case, GRSSE reduces to greedy rejection sampling \cite{harsha2010communication}. Nevertheless, using a nontrivial code with a larger $k$ can reduce the running time and the communication cost, since it can reduce the log term in (\ref{eq:grsse_comm}). For an AXN channel with bounded noise weight $\mathrm{wt}(\mathbf{Z})\le w$ (e.g., the Hamming ball channel), we should select a code with $k$ as large as possible while ensuring $d\ge2w+1$. 

Before we present the proof of Theorem \ref{thm:GRSSE_bound}, we introduce a lemma which may be of interest beyond GRSSE. Unlike greedy rejection sampling (GRS) \cite{harsha2010communication} which has uniquely defined and tractable iterations, the only information we know about the iterations in GRSSE is that $\kappa_{i}$ gives a low rejection probability $\mathbb{E}[\kappa_{i}(\mathrm{e}|\mathbf{S}_{i})]$ (or high acceptance probability) in (\ref{eq:grsse_argmin}) (the argmin might not even be unique). This flexibility poses a significant challenge to the analysis. The following lemma, which generalizes the bound for GRS in \cite{harsha2010communication}, shows that the expected logarithm of the number of iterations can be bounded simply by virtue of having a high acceptance probability for each iteration. It also improves upon \cite{harsha2010communication,flamich2023adaptive} by giving a smaller constant. The proof is in Appendix \ref{subsec:pf_gen_grs}.\smallskip{}

\begin{lem}
[General greedy rejection sampling lemma] \label{lem:gen_grs}Consider jointly-distributed random variables $Y\in\mathcal{Y}$ and $L\in\mathbb{N}$, a distribution $Q$ over $\mathcal{Y}$ with $P_{Y}\ll Q$, and $\theta>0$. If
\begin{align}
\mathbb{P}(L=i\,|\,L\ge i) & \ge\theta\int_{\mathcal{Y}}\min\left\{ \frac{\mathrm{d}P_{Y|L\ge i}}{\mathrm{d}Q}(y),\,1\right\} Q(\mathrm{d}y)\label{eq:gen_grs_ineq}
\end{align}
for every $i\in\mathbb{N}$ with $\mathbb{P}(L\ge i)>0$, then
\[
\mathbb{E}[\log L]\le D_{\mathrm{KL}}(P_{Y}\Vert Q)-\log\theta+(1+e^{-1})\log e.
\]
\end{lem}
\smallskip{}

We remark that for GRS \cite{harsha2010communication}, equality holds in (\ref{eq:gen_grs_ineq}) for $\theta=1$, so Lemma \ref{lem:gen_grs} applies to GRS. Lemma \ref{lem:gen_grs} generalizes GRS by allowing the event $L=i$ to be assigned arbitrarily to different values of $y$ (i.e., $\mathbb{P}(L=i|Y=y)$ can be arbitrary), as long as its probability satisfies the lower bound in (\ref{eq:gen_grs_ineq}). Lemma \ref{lem:gen_grs} also gives a smaller constant $(1+e^{-1})\log e\approx1.973$ compared to the constant $2\log e\approx2.885$ in \cite[proof of Claim IV.2]{harsha2010communication}, and the $\log(2e)\approx2.443$ in \cite{flamich2023adaptive}.\footnote{We remark that the Poisson functional representation (PFR) satisfies $\mathbb{E}[\log L]\le D_{\mathrm{KL}}(P_{Y}\Vert Q)+\log2$ \cite{li2024channel}, giving a smaller constant compared to GRS, though PFR is a rather different scheme (e.g., PFR is a noncausal sampling scheme \cite{liu2018rejection} whereas GRS is causal), and does not satisfy (\ref{eq:gen_grs_ineq}).}

We give a sketch of the proof of Theorem \ref{thm:GRSSE_bound}. The complete proof is in Appendix \ref{subsec:pf_GRSSE_bound}. We apply Lemma \ref{lem:gen_grs} on $(\mathbf{Z},L)$, $\theta=1$,
\[
Q(\mathbf{z})=\frac{1}{2}\mathbbm{1}\Big\{\mathrm{wt}(\mathbf{Z})<\frac{d}{2}\Big\} q^{k-n}+\frac{1}{2}q^{-n},
\]
which will yield the desired bounds.\footnote{Technically $\sum_{\mathbf{z}}Q(\mathbf{z})\neq1$. Refer to Appendix \ref{subsec:pf_GRSSE_bound} for the rigorous proof.} To check that the acceptance probability at iteration $i$ is at least $F:=\sum_{\mathbf{z}}\min\{p_{\mathbf{Z}}^{(i)}(\mathbf{z}),\,Q(\mathbf{z})\}$ as required by Lemma \ref{lem:gen_grs}, we need to find one particular $\kappa_{i}(\mathbf{z}|\mathbf{s})$ which gives an acceptance probability at least $F$. Given $\mathbf{s}$, this $\kappa_{i}$ either picks a $\mathbf{z}$ uniformly at random among $\mathbf{z}$'s satisfying $\mathbf{z}\mathbf{H}^{\top}=\mathbf{s}$ with probability $1/2$, or picks the coset leader (the $\mathbf{z}$ giving the smallest $\mathrm{wt}(\mathbf{z})$ with $\mathbf{z}\mathbf{H}^{\top}=\mathbf{s}$) with probability $1/2$. This $\kappa_{i}$ will ensure that a $\mathbf{z}$ with small $\mathrm{wt}(\mathbf{z})$ is selected with enough probability, and hence simulating the part of the noise distribution where the noise weight is small (usually the hardest part to simulate). Since a $\mathbf{z}$ with $\mathrm{wt}(\mathbf{Z})<d/2$ must be a coset leader, we can verify that the flow is indeed at least $F$. 

\smallskip{}

\section{Mixed GRSSE and Experiments }

The $\mathbf{H}$-GRSSE in Theorem \ref{thm:GRSSE_bound} uses $\mathbf{H}_{i}=\mathbf{H}$ at iteration $i$ which are all the same code. It may be beneficial to allow the inner code $\mathbf{H}_{i}$ to depend on the iteration. Recall that in GRSSE, at iteration $i$, the channel noise distribution to be simulated is $p_{\mathbf{Z}}^{(i)}(\mathbf{z})$, which may change with $i$. Intuitively, in earlier iterations, $p_{\mathbf{Z}}^{(i)}(\mathbf{z})$ is more spread out (i.e., the channel simulated is noisy) so we can use a code $\mathbf{H}_{i}$ with lower rate, or even the trivial code where $k_{i}=0$. In later iterations where $p_{\mathbf{Z}}^{(i)}(\mathbf{z})$ is concentrated around $\mathbf{z}$'s with $\mathrm{wt}(\mathbf{z})\approx0$ (i.e., we want to send $\mathbf{X}$ almost noiselessly), it is better to use a code with higher rate, or even the complete code where $k_{i}=n$ (i.e., simply send $\mathbf{M}=\mathbf{X}$).

The $\mathbf{H}^{[1]}/\cdots/\mathbf{H}^{[N]}$\emph{-GRSSE}, or the \emph{mixed GRSSE}, is the GRSSE where each $\mathbf{H}_{i}$ is selected within a prescribed sequence of codes $\mathbf{H}^{[1]},\ldots,\mathbf{H}^{[N]}$ in this order, i.e., $(\mathbf{H}_{i})_{i}$ is formed by repeating $\mathbf{H}^{[1]}$ a nonnegative integer number of times, then $\mathbf{H}^{[2]}$ a nonnegative integer number of times, etc. The precise number of times can be determined by a computer search that attempts to minimize the expected communication cost. A simple efficient rule that performs well in experiments is to simply select the code $\mathbf{H}^{[j_i]}$ at iteration $i$ via
\begin{equation}
j_i := \max \big\{j \in [N]:\, \mathbb{P}_{\mathbf{Z} \sim p_{\mathbf{Z}}^{(i)}}(\mathrm{wt}(\mathbf{Z}) \ge \tilde{d}_j/2) \le \epsilon \big\},\label{eq:mixed_GRSSE_rule}
\end{equation}
where $\epsilon>0$ is a small fixed parameter, and $\tilde{d}_j$ is the distance of the code $\mathbf{H}^{[j]}$ (or an estimate of the distance), assuming that $\tilde{d}_1>\cdots > \tilde{d}_N$. 
The reason is that Theorem~\ref{thm:GRSSE_bound} suggests that a smaller $\mathbb{P}(\mathrm{wt}(\mathbf{Z}) \ge d/2)$ would result in a smaller communication cost. 
This rule shares some similarities with the adaptive greedy rejection sampling scheme~\cite{flamich2023adaptive}, which also adaptively changes the proposal distribution at each iteration.

We discuss another reason to consider mixed GRSSE. A limitation of the $\mathbf{H}$-GRSSE is that the precomputation of $\kappa_{i}$ in Algorithm \ref{alg:GRSSE} may not terminate, i.e., the number of iterations is unbounded. Even though Algorithm \ref{alg:GRSSE} does not terminate, the actual encoding and decoding in Algorithm \ref{alg:RSSE} almost surely terminates since $\mathbb{E}[\log L]<\infty$ by Theorem \ref{thm:GRSSE_bound}. In practice, we can run Algorithm \ref{alg:GRSSE} in a ``lazy'' manner, where we only compute $\kappa_{i}$ when the encoding algorithm in Algorithm \ref{alg:RSSE} reaches iteration $i$. This way, we can ensure a finite running time for $\mathbf{H}$-GRSSE. Nevertheless, if we require the precomputation in Algorithm \ref{alg:GRSSE} to terminate in finite time (e.g., if we want to compute the expected communication cost precisely), we can consider the $\mathbf{H}/\text{complete}$-GRSSE, where $\mathbf{H}^{[1]}=\mathbf{H}$ and $\mathbf{H}^{[2]}$ is the empty matrix (the complete code). When the encoder reachs the iteration where $\mathbf{H}_{i}$ is the complete code, it will simply send $\mathbf{M}=\mathbf{X}$ and terminate the algorithm.

Figure \ref{fig:grsse_bsc24} considers the simulation of the memoryless binary symmetric channel with $n=24$ and $\alpha\in[0,1/2]$. The plotted curves are:
\begin{enumerate}
\item The Golay-GRSSE algorithm, which is the $\mathbf{H}$-GRSSE with inner code $\mathbf{H}$  being the  $[24,12,8]$ Golay code \cite{golay1949digitalcoding}. The expected communication rate (expected encoding length divided by $n$) is $(12+\ell)/n$, where $\ell$ is the expected length of the Huffman encoding of $L$, with $p_{L}$ being computed via Algorithm \ref{alg:GRSSE}.
\item The mixed-GRSSE algorithm using the rule \eqref{eq:mixed_GRSSE_rule}, with $\epsilon = 10^{-9}$ and $\mathbf{H}^{[1]},\ldots,\mathbf{H}^{[N]}$ being the best known $[24,k]$ linear block codes \cite{Grassl:codetables} with the largest distance for $k=0,1,2,3,5,7,12,14,18,19,23,24$, computed using Magma \cite{bosma1997magma}. 
\item Poisson functional representation (PFR) \cite{sfrl_trans}. Only the upper bound in \cite{li2024pointwise} is plotted since the blocklength $n=24$ is too large for PFR to be performed efficiently.
\item Greedy rejection sampling (GRS) \cite{harsha2010communication}. Only an upper bound computed via Lemma \ref{lem:gen_grs} (which improves upon \cite{harsha2010communication,flamich2023adaptive}) is plotted since the blocklength $n=24$ is too large for GRS to be performed efficiently.
\item The capacity of the channel, which is a lower bound on the communication rate.
\item The excess functional information lower bound in \cite[Proposition 1]{sfrl_trans}, which is a tighter lower bound for finite $n$. (Also see \cite{goc2024causal,flamich2025redundancy} for a generalization.)
\end{enumerate}

Since the Golay code has a distance $8$, according to Theorem~\ref{thm:GRSSE_bound}, Golay-GRSSE is only suitable when $\mathbb{P}(\mathrm{wt}(\mathbf{Z}) \le 3)$ is large, meaning that it requires $\alpha \le 3/24$. Practically, $\alpha$ has to be lower to ensure $\mathrm{wt}(\mathbf{Z}) \le 3$ most of the time. We can see from Figure \ref{fig:grsse_bsc24} that Golay-GRSSE is suitable for $\alpha\le0.1$, where its rate is close to the excess functional information lower bound. 
In comparison, the mixed-GRSSE is close to the lower bound for every $\alpha$, due to the ability to choose a code with a suitable distance at each iteration. The gap between the rate of mixed-GRSSE and the excess functional information lower bound is at most $0.06$.

Figure \ref{fig:grsse_hamming24} considers the simulation of the Hamming ball channel with $n=24$ and relative weights $\alpha=w/n$ for $w=0,\ldots,12$, which also serves as lossy source coding under Hamming distortion with almost-sure distortion constraint at distortion level $\alpha$. We plot the same simulation algorithms as Figure \ref{fig:grsse_bsc24}, and also the best known conventional linear covering codes \cite{cohen1985covering} (note that these codes do not simulate the Hamming ball channels; they merely ensure the same upper bound on the Hamming distortion as the Hamming ball channels). We can see that mixed-GRSSE can outperform conventional linear codes as lossy source codes for distortion between $4/24$ and $10/24$. Mixed-GRSSE also has the advantage of exact channel simulation---the error is exactly uniform over the Hamming ball. This is a similar phenomenon as rejection-sampled universal quantization \cite{ling2024rejection}, where channel simulation can also outperform conventional lattice codes in terms of rate-distortion performance.

\begin{figure*}
\begin{centering}
\includegraphics[scale=0.5]{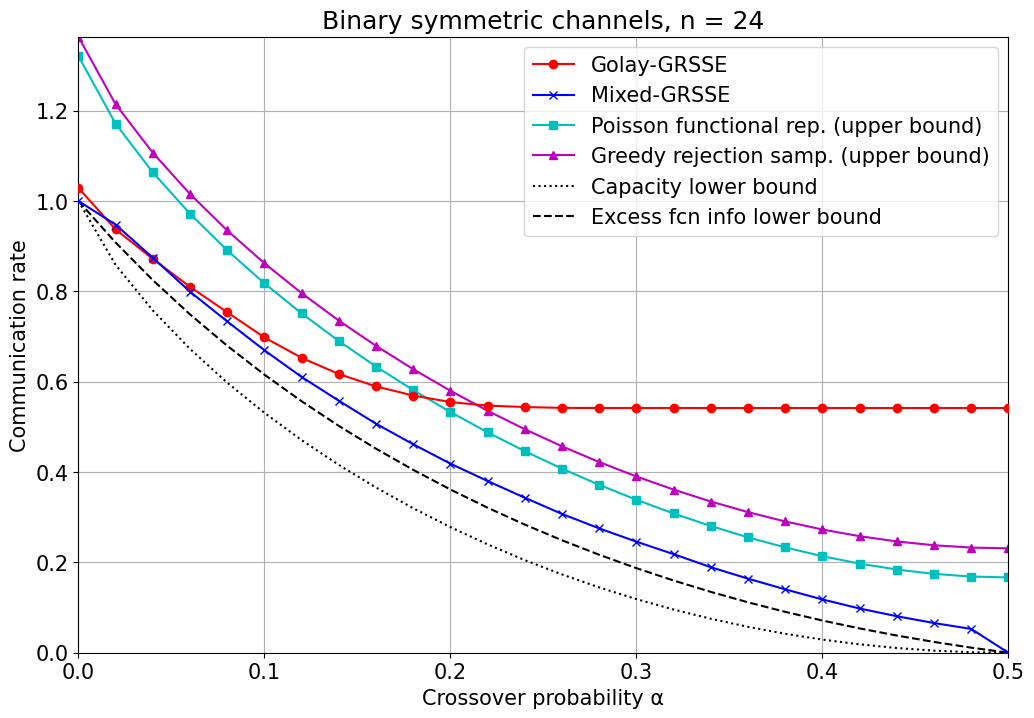}
\par\end{centering}
\caption{\label{fig:grsse_bsc24}The communication rate needed by various algorithms to simulate the memoryless binary symmetric channel with $n=24$ and $\alpha\in[0,1/2]$.}

\end{figure*}

\begin{figure*}
\begin{centering}
\includegraphics[scale=0.5]{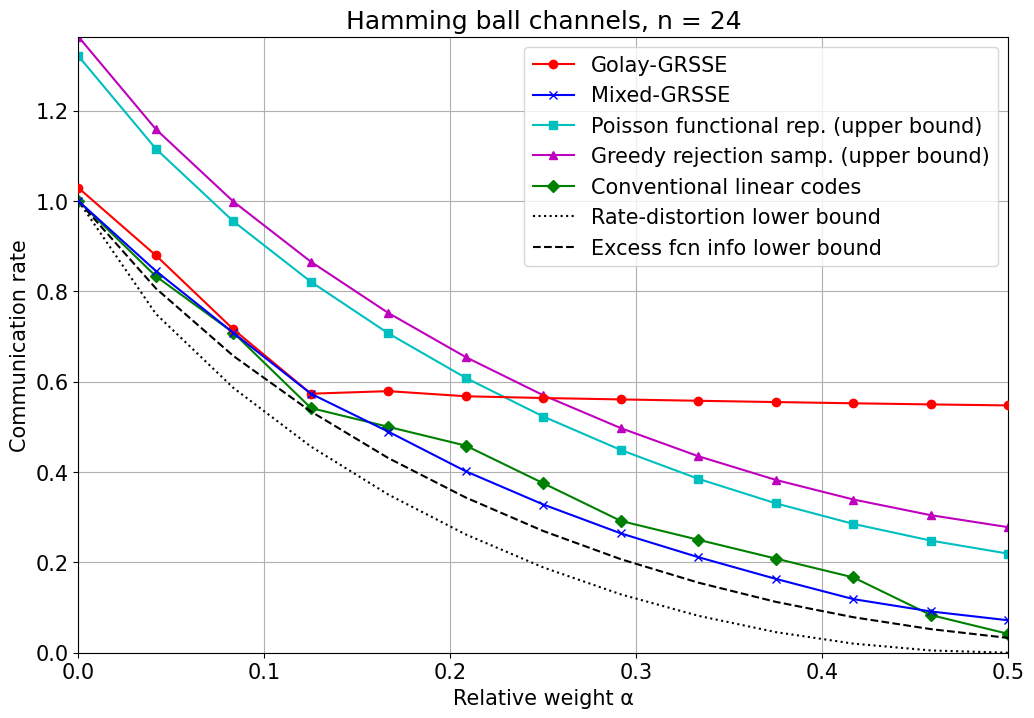}
\par\end{centering}
\caption{\label{fig:grsse_hamming24}The communication rate needed by various algorithms to simulate the Hamming ball channel with $n=24$ and relative weights $\alpha=w/n$ for $w=0,\ldots,12$.}
\end{figure*}

\section{GRSSE with Juxtaposition Codes}
We now discuss a method to apply GRSSE with a larger blocklength $n$. A simple way to increase the blocklength of a code is to perform juxtaposition, i.e., a codeword in the resultant code is the juxtaposition of any $r$ codewords of the original code. The parity-check matrix of the juxtaposition code is the block-diagonal matrix 
$\mathbf{I}_r \otimes \mathbf{H}$, where
where $\mathbf{H}$ is the parity-check matrix of the original code, $\mathbf{I}_r$ is the $r \times r$ identity matrix, and $\otimes$ is the Kronecker product. The type set distribution $p_{\mathcal{T}_{\mathbf{I}_r \otimes \mathbf{H}}}$ \eqref{eq:typeset_dist} of the juxtaposition code $\mathbf{I}_r \otimes \mathbf{H}$ can be computed efficiently in terms of the type set distribution $p_{\mathcal{T}_{\mathbf{H}}}$ of the original code, by noting that $p_{\mathcal{T}_{\mathbf{I}_r \otimes \mathbf{H}}}$ is the distribution of $\{r^{-1} \sum_{i=1}^r \mathbf{t}_i: \mathbf{t}_1 \in \mathcal{T}_1,\ldots, \mathbf{t}_r \in \mathcal{T}_r\}$ where $\mathcal{T}_i \stackrel{\mathrm{iid}}{\sim} p_{\mathcal{T}_{\mathbf{H}}}$, i.e., $p_{\mathcal{T}_{\mathbf{I}_r \otimes \mathbf{H}}}$ is the convolution (with respect to Minkowski sum) of $r$ copies of $p_{\mathcal{T}_{\mathbf{H}}}$.

Figure \ref{fig:grsse_bsc96} considers the simulation of the memoryless binary symmetric channel with $n=96$ (except PolarSim) and $\alpha\in[0,1/5]$. 
The plotted curves are:
\begin{enumerate}
\item The $4 \times$Golay-GRSSE with $n=96$, which is the pure GRSSE with inner code $\mathbf{I}_4 \otimes \mathbf{H}$ being the the juxtaposition of four $[24,12,8]$ Golay codes.\footnote{For the sake of efficiency, the number of iterations for both $\mathbf{I}_4 \otimes \mathbf{H}$-GRSSE and mixed-GRSSE is capped at $20000$, i.e., we switch to the complete code at iteration $20000$.}
\item The mixed-GRSSE with $n=96$ using the rule \eqref{eq:mixed_GRSSE_rule}, with $\epsilon = 0.01$, $\mathbf{H}^{(1)},\ldots,\mathbf{H}^{(N)}$ being the trivial code ($k=0$), the repetition code ($k=1$), and the $4$-fold juxtaposition of the best known $[24,k]$ linear block codes \cite{Grassl:codetables} with the largest distance for $k=12,14,18,19,23,24$, computed using Magma \cite{bosma1997magma}.\footnote{We choose the better one among this rule and fixing the code to be the $4$-fold juxtaposition of the Golay code (same as the $4 \times$Golay-GRSSE).} For the $r$-fold juxtaposition of a code with distance $d_j$, we use an estimated distance $\tilde{d}_j = rd_j$ in rule \eqref{eq:mixed_GRSSE_rule}.\footnote{Technically, the distance of the $r$-fold juxtaposition is still $d_j$. Nevertheless, there are only $r$ codewords with weight $d_j$, and their contribution is small in the GRSSE algorithm, which degrades gracefully when a small number of codewords have weights that are too small. Intuitively, the ``effective distance'' of the $r$-fold juxtaposition is $\tilde{d}_j = rd_j$, and this estimate performs well in the experiment.}
\item PolarSim (channel simulation via polar codes) \cite{sriramu2024fast} with $n = 4096$.
\item Poisson functional representation (PFR) \cite{sfrl_trans} and greedy rejection sampling (GRS) \cite{harsha2010communication} with $n=96$. Only upper bounds (\cite{li2024pointwise} and Lemma \ref{lem:gen_grs}) are plotted since the blocklength $n=96$ is exceedingly large for these two schemes to be performed within a year's time. They are only the hypothetical performance of the impractical schemes, and should not be compared directly to GRSSE which is efficient.
\item The lower bounds given by the capacity of the channel, and the excess functional information lower bound in \cite[Proposition 1]{sfrl_trans} with $n=96$.
\end{enumerate}

Figure \ref{fig:grsse_bsc96}  shows that $4 \times$Golay-GRSSE performs well when $\alpha \in [0.1, 0.125]$.
We can also see that mixed-GRSSE performs well when $\alpha \le 0.125$.
PolarSim achieves a smaller communication rate, but requires a significantly larger $n$ to have a satisfactory performance. 
The difference between the rates of mixed-GRSSE ($n=96$) and PolarSim ($n = 4096$) is small when $\alpha \le 0.125$ (the difference is at most $0.058$). 

With GRSSE, a scheme with small blocklength but a comparable performance with PolarSim which uses a significantly larger blocklength, we show that a large blocklength is unnecessary to achieve an almost-optimal performance. The applicability of GRSSE under small blocklength allows channel simulation to be performed on shorter sequences with smaller delay.

\begin{figure*}
\begin{centering}
\includegraphics[scale=0.5]{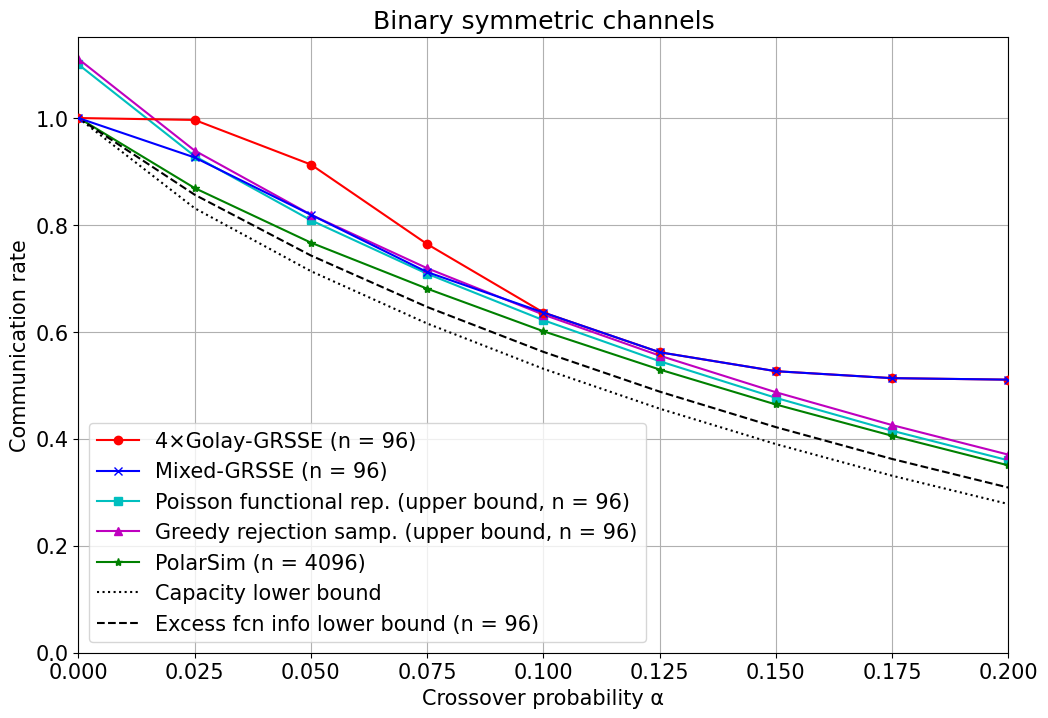}
\par\end{centering}
\caption{\label{fig:grsse_bsc96}The communication rate needed by various algorithms to simulate the memoryless binary symmetric channel with $n=96$ (except PolarSim) and $\alpha\in[0,1/5]$. The data for the plot for PolarSim is extracted from \cite[Figure 5]{sriramu2024fast}.}

\end{figure*}

\section{Acknowledgement}

This work was partially supported by two grants from the Research Grants Council of the Hong Kong Special Administrative Region, China {[}Project No.s: CUHK 24205621 (ECS), CUHK 14209823 (GRF){]}.

\smallskip{}

\fi

\bibliographystyle{IEEEtran}
\bibliography{ref}

\iffullver

\appendix{}

\subsection{Proof of Theorem \ref{thm:polar-RSSE}\label{pf:thm_polar_RSSE}}


It is evident that the sequences of $(n, 1)$ and $(n, 0)$-polar RSSEs satisfy the statement of Theorem \ref{thm:polar-RSSE} for $D = 0$ and $1/2$, respectively.

We now consider a fixed $D \in (0, 1/2)$. Let $\mathcal{Q}^{(n)}$ be the distribution of the quantization noise generated by $(\mathbf{H}^{(n)}, \kappa^{(n)})$, defined as
\begin{equation*}
    \mathcal{Q}^{(n)}(\mathbf{z}) = \mathbb{E}[\kappa^{(n)}(\mathbf{z}|\mathbf{X}\mathbf{H}^{(n)\top})]
\end{equation*}
for all $\mathbf{z} \in \mathbb{F}_2^n$, where $\mathbf{X}\sim \mathrm{Unif}(\mathbb{F}_2^n)$.
Let $p_{\mathrm{Bern}(D)}^{\otimes n}$ denote the distribution of $n$ i.i.d. Bernoulli variables with parameter $D$.
Building on Lemma 11, Equation (22) and Theorem 19 of \cite{koradaPolarCodesAre2010}, we obtain Corollary \ref{thm:polar_noise_dist}.

\begin{cor} \label{thm:polar_noise_dist}
Fix a target distortion $D \in (0, 1/2)$.
For any $R > 1 - H_\mathrm{b}(D)$ and any $\beta \in (0, 1/2)$, there exists a sequence of polar codes of length $n$ with rates $R^{(n)} < R$ such that under SC encoding using randomized rounding the distribution $\mathcal{Q}^{(n)}$ of their quantization noise satisfies
\begin{equation*}
    \sum_{\mathbf{z}\in\mathbb{F}_2^n} \Big| \mathcal{Q}^{(n)}(\mathbf{z}) - p_{\mathrm{Bern}(D)}^{\otimes n}(\mathbf{z}) \Big| = O(2^{-n^\beta}).
\end{equation*}
\end{cor}

Consider a sequence of polar codes satisfying the statement of Corollary \ref{thm:polar_noise_dist}.
Define 
\begin{equation*}
    Q^{(n)}(w) := \sum_{\mathbf{z} \in \mathbb{F}_2^n\{w\}} \mathcal{Q}^{(n)}(\mathbf{z}).
\end{equation*}
for all $w \in [0:n]$.
Based on Lemma \ref{thm:ref_distr} and Corollary \ref{thm:RS_P_Q} (also see Example \ref{eg:RTC}), the $(n, R^{(n)})$-polar-RSSE for simulating $\mathrm{CWC}^{(n)}(w^{(n)})$ achieves a communication rate of at most
\begin{equation*}
    R^{(n)} + \frac{1}{n} \big(\log e + 1 -\log Q^{(n)}(w^{(n)})\big)
\end{equation*}
with expected $Q^{(n)}(w^{(n)}){}^{-1}$ SC encoding iterations.
Thus, it suffices to show $Q^{(n)}(w^{(n)}) = \Omega(1/\sqrt{n})$.

Let $W^{(n)}$ be a binomial variable with parameters $n$ and $D$.
Then we have 
\begin{align}
    &~\Big| Q^{(n)}(w^{(n)}) - p_{W^{(n)}}(w^{(n)}) \Big| \notag \\ \displaybreak[0]
    \leq& \sum_{w=0}^{n} \Big| Q^{(n)}(w) - p_{W^{(n)}}(w) \Big| \notag \\ \displaybreak[0]
    \overset{\text{(a)}}{=}& \sum_{w=0}^{n} \Big| \sum_{\mathbf{z}\in\mathbb{F}_2^n\{w\}} \big(\mathcal{Q}^{(n)}(\mathbf{z}) - p_{\mathrm{Bern}(D)}^{\otimes n}(\mathbf{z})\big) \Big| \notag \\ \displaybreak[0]
    \overset{\text{(b)}}{\leq}& \sum_{w=0}^{n} \sum_{\mathbf{z}\in\mathbb{F}_2^n\{w\}} \Big| \mathcal{Q}^{(n)}(\mathbf{z}) - p_{\mathrm{Bern}(D)}^{\otimes n}(\mathbf{z}) \Big| \notag \\ \displaybreak[0]
    =& \sum_{\mathbf{z}\in\mathbb{F}_2^n} \Big| \mathcal{Q}^{(n)}(\mathbf{z}) - p_{\mathrm{Bern}(D)}^{\otimes n}(\mathbf{z}) \Big| \notag \\ \displaybreak[0]
    \overset{\text{(c)}}{=}&~ O(2^{-n^\beta}) \label{eq:proof-Q_P_bound}
\end{align}
for some $\beta \in (0, 1/2)$, 
where (a) holds because $W^{(n)}$ is the sum of $n$ i.i.d. Bernoulli variables with parameter $D$; 
(b) follows from the triangle inequality;
and (c) is given by Corollary \ref{thm:polar_noise_dist}.
Moreover, based on the local limit theorem, as $n \to \infty$, we have 
\begin{align} \label{eq:proof-LLT}
    &\sup_{w \in [0:n]} \Bigg|\sqrt{D(1-D)n} p_{W^{(n)}}(w) - \frac{\exp\left(\frac{-(w-nD)^2}{2D(1-D)n}\right)}{\sqrt{2\pi}}\Bigg| \to 0.
\end{align}
Since $|w^{(n)} - nD| = O(\sqrt{n})$, there exist $c_1 > 0$ and $N_1 \in \mathbb{N}$ such that $(w^{(n)} - nD)^2 \leq c_1^2 n$ and
\begin{equation*}
    \frac{1}{\sqrt{2\pi}}\exp\left(\frac{-(w^{(n)}-nD)^2}{2D(1-D)n}\right) \geq \underbrace{\frac{1}{\sqrt{2\pi}}\exp\left(\frac{-c_1^2}{2D(1-D)}\right)}_{\delta_1}
\end{equation*}
for all $n \geq N_1$.
According to \eqref{eq:proof-LLT}, there exists $N_2 \in \mathbb{N}$ such that 
\begin{align} \label{eq:proof-P_w_bound}
&\Bigg|\sqrt{D(1-D)n} p_{W^{(n)}}(w^{(n)}) - \frac{\exp\left(\frac{-(w^{(n)}-nD)^2}{2D(1-D)n}\right)}{\sqrt{2\pi}}\Bigg| \leq \frac{\delta_1}{2}
\end{align}
holds for all $n \geq N_2$.
Therefore, for all $n \geq \max\{N_1, N_2\}$, 
\begin{align*}
    &~ \sqrt{D(1-D)n} p_{W^{(n)}}(w^{(n)}) \\
    \geq &~ \frac{1}{\sqrt{2\pi}}\exp\left(\frac{-(w^{(n)}-nD)^2}{2D(1-D)n}\right) - \frac{1}{2}\delta_1 \\
    \geq &~ \delta_1 - \frac{1}{2}\delta_1 \\
    = &~ \frac{1}{2}\delta_1,
\end{align*}
and thus
\begin{equation*}
    p_{W^{(n)}}(w^{(n)}) \geq \frac{\delta_1}{2\sqrt{D(1-D)n}},
\end{equation*}
i.e., $p_{W^{(n)}}(w^{(n)}) = \Omega(1/\sqrt{n})$.
In conjunction with \eqref{eq:proof-Q_P_bound}, it can be seen that $Q^{(n)}(w^{(n)}) = \Omega(1/\sqrt{n})$.

\subsection{Proof of Lemma \ref{lem:gen_grs}\label{subsec:pf_gen_grs}}

Let $S_{L}(i):=\mathbb{P}(L\ge i)$, $S_{L|Y}(i|y):=\mathbb{P}(L\ge i|Y=y)$. By (\ref{eq:gen_grs_ineq}) and Bayes' rule,
\begin{align*}
 & \mathbb{P}(L=i)\\
 & \ge\theta S_{L}(i)\int_{\mathcal{Y}}\min\left\{ \frac{\mathrm{d}P_{Y|L\ge i}}{\mathrm{d}Q}(y),\,1\right\} Q(\mathrm{d}y)\\
 & =\theta\int_{\mathcal{Y}}\min\left\{ \frac{S_{L|Y}(i|y)}{S_{L}(i)}\frac{\mathrm{d}P_{Y}}{\mathrm{d}Q}(y),\,1\right\} S_{L}(i)Q(\mathrm{d}y)\\
 & \ge\theta\int_{\mathcal{Y}}\min\left\{ \frac{\mathrm{d}P_{Y}}{\mathrm{d}Q}(y),\,1\right\} S_{L}(i)S_{L|Y}(i|y)Q(\mathrm{d}y).
\end{align*}
Hence,
\begin{align*}
 & 1=\sum_{i=1}^{\infty}\mathbb{P}(L=i)\\
 & \ge\theta\sum_{i=1}^{\infty}\int_{\mathcal{Y}}\min\left\{ \frac{\mathrm{d}P_{Y}}{\mathrm{d}Q}(y),\,1\right\} S_{L}(i)S_{L|Y}(i|y)Q(\mathrm{d}y)\\
 & \ge\theta\int_{\mathcal{Y}}\min\left\{ \frac{\mathrm{d}P_{Y}}{\mathrm{d}Q}(y),\,1\right\} \\
 & \;\;\cdot\left(\sum_{i=1}^{\infty}\left(iS_{L}(i)-(i-1)S_{L}(i-1)\right)S_{L|Y}(i|y)\right)Q(\mathrm{d}y)\\
 & \ge\theta\int_{\mathcal{Y}}\min\left\{ \frac{\mathrm{d}P_{Y}}{\mathrm{d}Q}(y),\,1\right\} \mathbb{E}\left[LS_{L}(L)\,\big|\,Y=y\right]Q(\mathrm{d}y)\\
 & =\theta\int_{\mathcal{Y}}\min\left\{ \bigg(\frac{\mathrm{d}P_{Y}}{\mathrm{d}Q}(y)\bigg)^{-1},1\right\} \mathbb{E}\left[LS_{L}(L)\big|Y=y\right]P_{Y}(\mathrm{d}y)\\
 & =\theta\mathbb{E}\left[\min\left\{ \bigg(\frac{\mathrm{d}P_{Y}}{\mathrm{d}Q}(y)\bigg)^{-1},\,1\right\} LS_{L}(L)\right]\\
 & =:\theta c.
\end{align*}
Define a probability measure $\tilde{Q}\ll P_{Y,L}$ over $\mathcal{Y}\times\mathbb{N}$, with
\[
\frac{\mathrm{d}\tilde{Q}}{\mathrm{d}P_{Y,L}}(y,i)=\frac{1}{c}\min\left\{ \bigg(\frac{\mathrm{d}P_{Y}}{\mathrm{d}Q}(y)\bigg)^{-1},1\right\} iS_{L}(i).
\]
Note that we also have $P_{Y,L}\ll\tilde{Q}$ since the above is positive $P_{Y,L}$-almost everywhere as $P_{Y}\ll Q$. Since $(Y,L)\sim P_{Y,L}$ and $c\le1/\theta$,
\begin{align*}
0 & \le D_{\mathrm{KL}}(P_{Y,L}\Vert\tilde{Q})\\
 & =\mathbb{E}\left[\log\frac{\mathrm{d}P_{Y,L}}{\mathrm{d}\tilde{Q}}(Y,L)\right]\\
 & \le\mathbb{E}\left[\log\max\left\{ \frac{\mathrm{d}P_{Y}}{\mathrm{d}Q}(Y),1\right\} \right]-\mathbb{E}[\log(\theta LS_{L}(L))].
\end{align*}
By \cite[proof of Claim IV.2]{harsha2010communication} (also see \cite[proof of Theorem 7]{li2024channel}), we have $\mathbb{E}[-\log S_{L}(L)]\le\log e$ for every random variable $L\in\mathbb{N}$. Also, by \cite[Appendix A]{harsha2007communication},
\[
\mathbb{E}\left[\log\max\left\{ \frac{\mathrm{d}P_{Y}}{\mathrm{d}Q}(Y),1\right\} \right]\le D_{\mathrm{KL}}(P_{Y}\Vert Q)+e^{-1}\log e.
\]
The result follows.

\subsection{Proof of Theorem \ref{thm:GRSSE_bound}\label{subsec:pf_GRSSE_bound}}

Let
\[
\mathbf{z}^{*}(\mathbf{s}):=\mathrm{argmin}_{\mathbf{z}\in\mathbb{F}_{q}^{n}:\,\mathbf{z}\mathbf{H}^{\top}=\mathbf{s}}\;\mathrm{wt}(\mathbf{z})
\]
be the coset leader.  Fix any $0\le\beta\le1$. Define a conditional distribution $\tilde{\kappa}(\mathbf{z}|\mathbf{s})$ for $\mathbf{z}\in\mathbb{F}_{q}^{n}$, $\mathbf{s}\in\mathbb{F}_{q}^{n-k}$ with 
\[
\tilde{\kappa}(\mathbf{z}|\mathbf{s})=\beta\mathbbm{1}\{\mathbf{z}=\mathbf{z}^{*}(\mathbf{s})\}+(1-\beta)q^{-k}\mathbbm{1}\{\mathbf{z}\mathbf{H}^{\top}=\mathbf{s}\},
\]
i.e., given the syndrome $\mathbf{s}$, we select the coset leader $\mathbf{z}^{*}(\mathbf{s})$ with probability $\beta$, or select an arbitrary $\mathbf{z}$ in the coset with probability $1-\beta$. Although $\tilde{\kappa}$ may not satisfy the constraint in (\ref{eq:grsse_argmin}), we can modify it to 
\[
\hat{\kappa}_{i}(\mathbf{z}|\mathbf{s})=\tilde{\kappa}(\mathbf{z}|\mathbf{s})\min\left\{ \frac{p_{\mathbf{Z}}^{(i)}(\mathbf{z})}{\mathbb{E}[\tilde{\kappa}(\mathbf{z}\boldsymbol{\Pi}_{i}\,|\,\mathbf{S}_{i})]},\,1\right\} ,
\]
$\hat{\kappa}_{i}(\mathrm{e}|\mathbf{s})=1-\sum_{\mathbf{z}}\hat{\kappa}_{i}(\mathbf{z}|\mathbf{s})$, which ensures that the constraint in (\ref{eq:grsse_argmin}) is satisfied. If GRSSE selects $\hat{\kappa}_{i}$, the proability of acceptance at iteration $i$ is
\begin{align*}
 & 1-\mathbb{E}[\hat{\kappa}_{i}(\mathrm{e}\,|\,\mathbf{S}_{i})]\\
 & =\mathbb{E}\bigg[\sum_{\mathbf{z}\in\mathbb{F}_{q}^{n}}\hat{\kappa}_{i}(\mathbf{z}\boldsymbol{\Pi}_{i}|\mathbf{S}_{i})\bigg]\\
 & =\mathbb{E}\bigg[\sum_{\mathbf{z}}\tilde{\kappa}(\mathbf{z}\boldsymbol{\Pi}_{i}|\mathbf{S}_{i})\min\bigg\{\frac{p_{\mathbf{Z}}^{(i)}(\mathbf{z})}{\mathbb{E}[\tilde{\kappa}(\mathbf{z}\boldsymbol{\Pi}_{i}\,|\,\mathbf{S}_{i})]},\,1\bigg\}\bigg]\\
 & =\sum_{\mathbf{z}}\min\left\{ p_{\mathbf{Z}}^{(i)}(\mathbf{z}),\,\mathbb{E}[\tilde{\kappa}(\mathbf{z}\boldsymbol{\Pi}_{i}\,|\,\mathbf{S}_{i})]\right\} \\
 & =\sum_{\mathbf{z}}\min\left\{ p_{\mathbf{Z}}^{(i)}(\mathbf{z}),\,\beta\mathbb{P}(\mathbf{z}\boldsymbol{\Pi}_{i}=\mathbf{z}^{*}(\mathbf{S}_{i}))+(1-\beta)q^{-n}\right\} \\
 & =:\sum_{\mathbf{z}}\min\left\{ p_{\mathbf{Z}}^{(i)}(\mathbf{z}),\,Q(\mathbf{z})\right\} .
\end{align*}
Since GRSSE selects the $\kappa_{i}$ with the maximum probability of acceptance, $\kappa_{i}$ is at least as good as the above. Hence, applying Lemma \ref{lem:gen_grs} on $(\mathbf{Z},L)$, $Q(\mathbf{z})$ and $\theta=1$ gives
\begin{align*}
 & \mathbb{E}\left[\log L\right]-\eta\\
 & \le D_{\mathrm{KL}}(p_{\mathbf{Z}}\Vert Q)\\
 & =-\mathbb{E}\left[\log Q(\mathbf{Z})\right]-H(\mathbf{Z})\\
 & =-\mathbb{E}\left[\log\left(\beta\mathbb{P}(\mathbf{Z}\boldsymbol{\Pi}_{i}=\mathbf{z}^{*}(\mathbf{S}_{i}))+(1-\beta)q^{-n}\right)\right]-H(\mathbf{Z})\\
 & \stackrel{(a)}{\le}-\mathbb{E}\!\left[\log\!\left(\!\beta\mathbbm{1}\Big\{\mathrm{wt}(\mathbf{Z})\!<\!\frac{d}{2}\Big\} q^{k-n}+(1\!-\!\beta)q^{-n}\!\right)\!\right]-H(\mathbf{Z})\\
 & =\log\frac{1}{(1-\beta)q^{-n}}\\
 & \;\;\;+\mathbb{P}(\mathrm{wt}(\mathbf{Z})<d/2)\log\frac{(1-\beta)q^{-n}}{\beta q^{-(n-k)}+(1-\beta)q^{-n}}-H(\mathbf{Z})\\
 & \le\log q^{n}-H(\mathbf{Z})+\log\frac{1}{1-\beta}\\
 & \;\;\;+\mathbb{P}(\mathrm{wt}(\mathbf{Z})<d/2)\log\frac{(1-\beta)q^{-n}}{\beta q^{-(n-k)}}\\
 & =C-\mathbb{P}(\mathrm{wt}(\mathbf{Z})<d/2)\bigg(k\log q+\log\frac{\beta}{1-\beta}\bigg)+\log\frac{1}{1-\beta},
\end{align*}
where (a) is because if $\mathrm{wt}(\mathbf{z})<d/2$, then $\mathbf{z}\boldsymbol{\Pi}_{i}$ is the coset leader for one out of $q^{n-k}$ cosets. Therefore, we obtain (\ref{eq:grsse_logl}) by taking $\beta=1/2$. To obtain (\ref{eq:grsse_comm}), applying the inequality $H(L)\le\mathbb{E}[\log L]+\log(\mathbb{E}[\log L]+1)+1$ in \cite{sfrl_trans}, if we apply Huffman coding on $L$, the expected encoding length of $(L,\mathbf{M})$ is at most
\begin{align*}
 & H(L)+\log|\mathbb{F}_{q}^{k}|+1\\
 & \le C-\mathbb{P}(\mathrm{wt}(\mathbf{Z})<d/2)k\log q+\eta+3\\
 & \;\;+\log\left(C-\mathbb{P}(\mathrm{wt}(\mathbf{Z})<d/2)k\log q+\eta+2\right)+k\log q,
\end{align*}
which is the desired result.

\fi

\end{document}